\documentclass[10pt,twocolumn,letterpaper]{article}

\usepackage{cvpr}              % To produce the CAMERA-READY version
\definecolor{cvprblue}{rgb}{0.21,0.49,0.74}
\usepackage[pagebackref,breaklinks,colorlinks,allcolors=cvprblue]{hyperref}
\usepackage{amsmath}
\usepackage[symbol]{footmisc}
\usepackage{lineno}

\DeclareMathOperator*{\argmin}{arg\,min}

\def\paperID{7666} % *** Enter the Paper ID here
\def\confName{ICCV}
\def\confYear{2025}

\begin{document}
\title{Text2VDM: Text to Vector Displacement Maps\\for Expressive and Interactive 3D Sculpting}

%%%%%%%%% AUTHORS - PLEASE UPDATE
\author{
% Hengyu Meng\\
% \and
Hengyu Meng$^{1}$ \qquad
Duotun Wang$^{1}$ \qquad
Zhijing Shao$^{1}$ \qquad
% \\
Ligang Liu$^{2}$ \qquad
Zeyu Wang$^{1,3}$\footnotemark[1] \qquad 
\\
$^{1}$The Hong Kong University of Science and Technology (Guangzhou)
\\
$^{2}$University of Science and Technology of China
\\
$^{3}$The Hong Kong University of Science and Technology
}

\twocolumn[{
\maketitle
\begin{center}
    \captionsetup{type=figure}
    \includegraphics[width=0.97\textwidth]{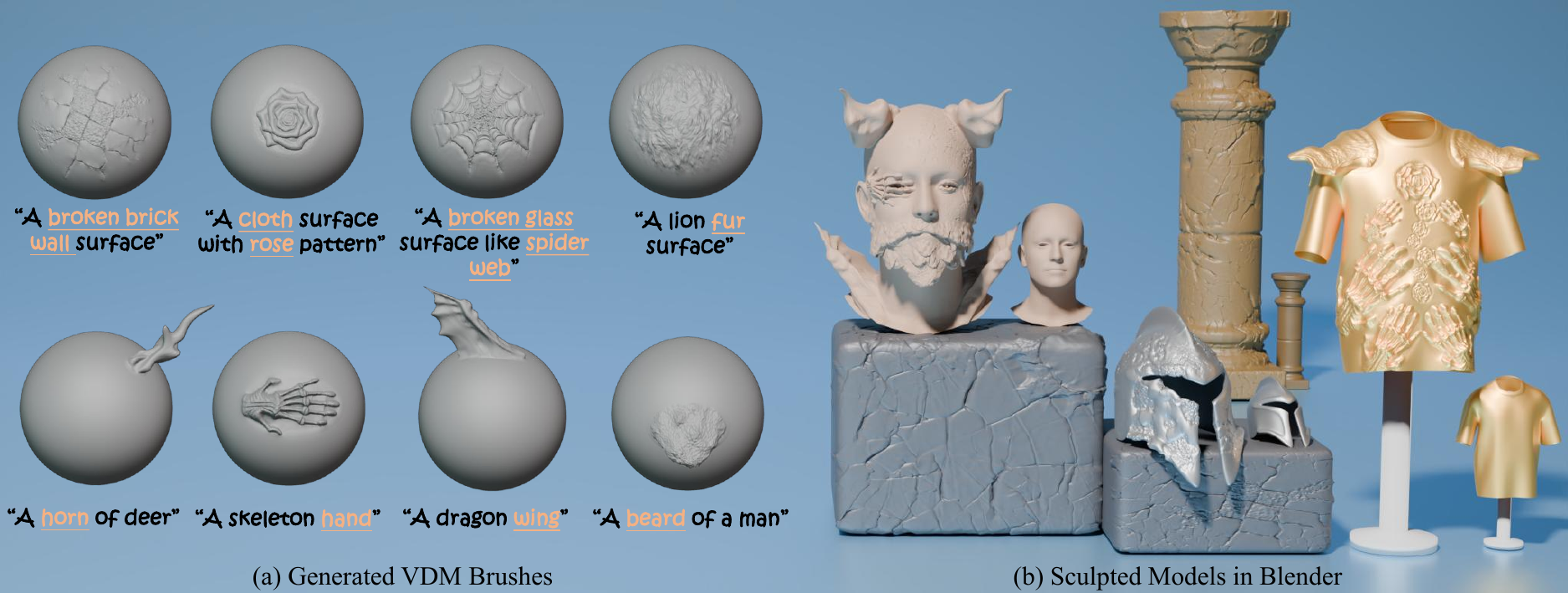}
    \captionof{figure}{
        \textbf{Example VDM brushes generated by Text2VDM and sculpted models in Blender.} Text2VDM can produce high-quality brushes for surface details (top row) and geometric structures (bottom row) from text input. Users can rapidly create an expressive model from a plain shape by directly applying these brushes in Blender. Yellow underlined text highlights semantics enhanced by our framework.
    }
    \label{Fig:teaser}
\end{center}
}]

\footnotetext[1]{Corresponding author.}

\begin{abstract}
% Zeyu comment 1107: 目前我改好的这版尽量做到了准确描述 descriptive，但是感觉还没有突出研究的动机（为什么重要、为什么难、为什么现有方法不好解决、我们的核心技术创新点在哪里）。请过一遍我的edits后再改一轮，注意逻辑的连贯性和sharpness。
% Professional 3D asset creation is an iterative and engaging process using sculpting brushes within modeling software. 
% Professional 3D asset creations require dynamic and iterative designs, with sculpting brushes serving as essential tools for precision and creativity. 
Professional 3D asset creation often requires diverse sculpting brushes to add surface details and geometric structures.
Despite recent progress in 3D generation, producing reusable sculpting brushes compatible with artists' workflows remains an open and challenging problem.
These sculpting brushes are typically represented as vector displacement maps (VDMs), which existing models cannot easily generate compared to natural images.
This paper presents Text2VDM, a novel framework for text-to-VDM brush generation through the deformation of a dense planar mesh guided by score distillation sampling (SDS).
The original SDS loss is designed for generating full objects and struggles with generating desirable sub-object structures from scratch in brush generation.
We refer to this issue as semantic coupling, which we address by introducing weighted blending of prompt tokens to SDS, resulting in a more accurate target distribution and semantic guidance.
Experiments demonstrate that Text2VDM can generate diverse, high-quality VDM brushes for sculpting surface details and geometric structures.
Our generated brushes can be seamlessly integrated into mainstream modeling software, enabling various applications such as mesh stylization and real-time interactive modeling.

% and enable repeatedly applying the geometry pattern of surface details or 3D components 

\end{abstract}

\section{Introduction}
\label{sec:intro}

Sculpting brushes are essential tools in 3D asset creation, as artists often require a variety of brushes to create surface details and geometric structures. In modeling software, 3D sculpting brushes are typically defined as vector displacement maps (VDMs). A VDM is a 2D image where each pixel stores a 3D displacement vector. Through these vectors, VDM brushes can create complex surface details, such as cracks and wood grain, or generate geometric structures like ears and horns. This allows artists to apply the same geometric pattern iteratively while sculpting.

\begin{figure}[tbh]
\centering
\includegraphics[width=0.46\textwidth]{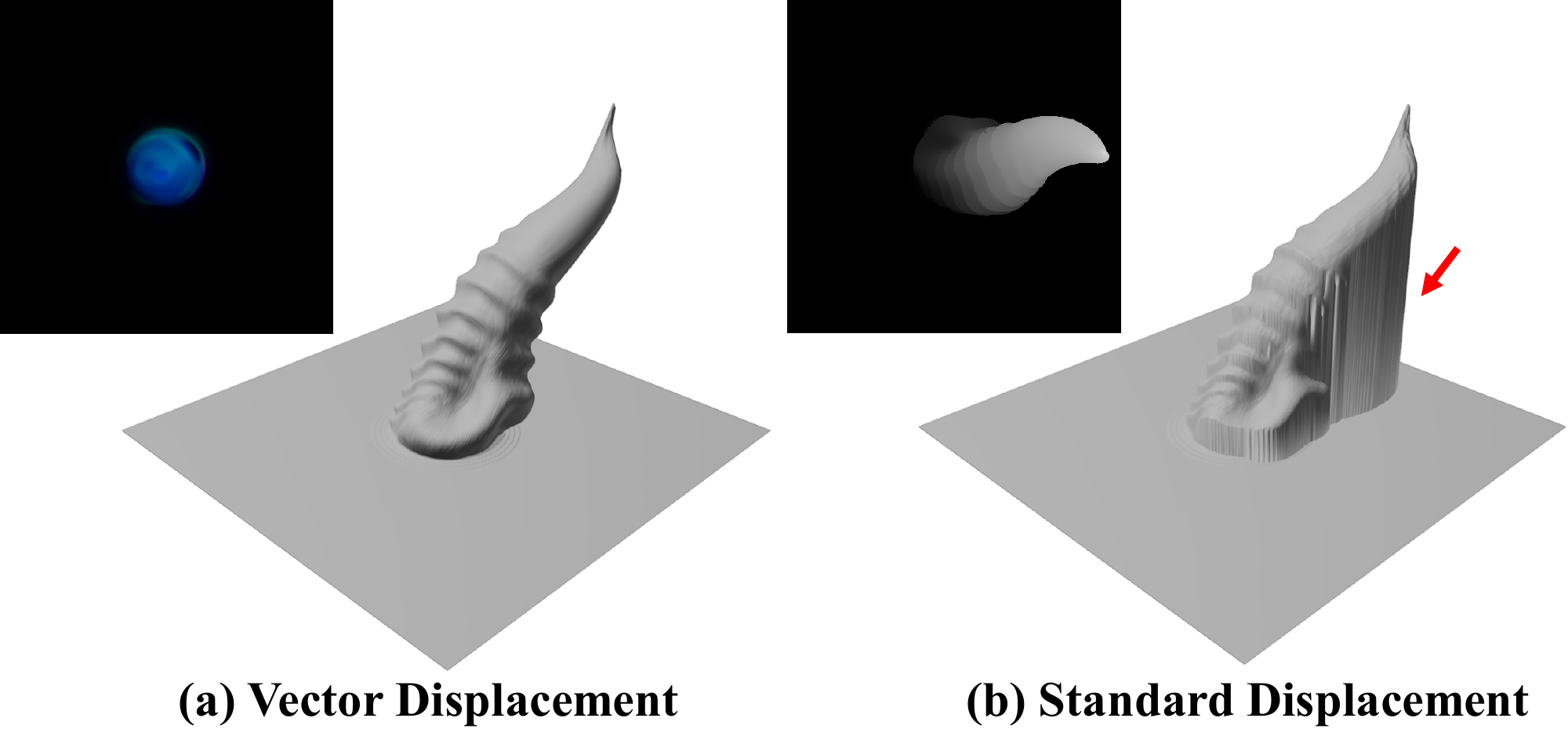}
\caption{
\textbf{Difference between vector and standard displacement.} The VDM enables full 3D vector displacement, while the height map only allows unidirectional standard displacement.}
\label{Fig: vector or height}
\end{figure}

Despite significant advances in text-to-image (T2I)~\cite{StableDiffusion:Arxiv:2021,dalle} and text-to-3D generation~\cite{Make3D:2023:ICCV,single1-instant3d,single2-wonder3d,DreamGaussian:Arxiv:2023,DreamFusion:ICLR:2022}, existing methods are unsuitable for creating VDM brushes. We summarize the challenges as follows: 1)~Since VDMs are not natural images (\Cref{Fig: vector or height}), it is difficult for existing T2I models to generate them directly. 2)~From a 3D perspective, a VDM represents mesh deformation through per-vertex displacement vectors from a dense planar mesh. Mapping any generated mesh to a dense planar mesh to create a VDM is non-trivial. 3)~Sculpting brushes often involve sub-object structures, whereas most 3D generation methods can only generate full objects. Enabling users to accurately control the generation of sculpting brushes through text prompts in a semantically focused manner remains challenging.

To address the challenges of brush generation, we propose Text2VDM, a novel optimization-based framework that generates diverse and controllable VDM brushes from text input. Our approach does not generate VDMs directly from a T2I model. Instead, we address VDM brush generation from a 3D perspective by applying score distillation with a pre-trained T2I model to guide mesh deformation.
% we formulate VDM brush generation as the geometry deformation from a dense planar mesh.
% to establish a precise mapping between the dense planar mesh and the generated mesh.  
% We design a mesh deformation framework to establish a precise mapping between the dense planar mesh and the generated mesh. 
Our framework supports three ways to initialize a base mesh through a zero-valued, spike-pattern, or user-specified VDM for custom shape control.
% a $512\times512$-resolution 3-channel VDM representing a dense planar mesh
% , where the three values of each pixel represent the three-axis displacement of the corresponding vertex.
% After constructing a dense planar mesh from the VDM, we provide a shape control method that allows users to use either the default spike-pattern VDM or a user-specified VDM to control the initial volume of the mesh.
% We reparameterize the mesh vertices through an implicit formulation based on the Laplace-Beltrami operator~\cite{Largesteps:SIGGRAPH:2021} to achieve high-quality optimization of mesh deformations.
For mesh deformation, we formulate a Sobolev preconditioned optimization~\cite{Largesteps:SIGGRAPH:2021} to maintain mesh quality with intrinsic smoothness.
We also provide optional region control using a mask of activated mesh deformation, helping users obtain the intended brush effects. The normal maps of the mesh are then rasterized by a differentiable renderer for brush optimization.

We observed that the standard score distillation sampling (SDS)~\cite{DreamFusion:ICLR:2022} can lead to semantic coupling 
% and fails to provide precise guidance 
when supervising the generation of sub-object level structures due to the associated semantics caused by the noisy gradients from the full object. For example, a generated deer's horn should not be a full deer's head, or a generated beard should not include a nose. A straightforward solution is to use negative prompts~\cite{CSD:Arxiv:2023, NFSD:Arxiv:2023} to exclude undesired semantics, but our experiments show that this semantics suppression approach is ineffective in decoupling semantics and leads to an unstable optimization process. Instead, we propose to enhance the semantics of part-related words by applying weighted blending to the tokens in the prompt.  This results in semantically focused text embedding, directing toward a more precise target distribution while reducing noisy gradients during optimization. 
% As a result, the generation process of our approach is stable and semantically focused. 

% To ensure the final brush effect aligns with user expectations, we provide two control methods: shape control and region control. Shape control uses a user-specified VDM to set the desired brush volume and direction, while region control employs a mask to define specific areas of mesh deformation.
% To better align the brush with user application requirements, we provide two control methods for the brush generation process: a shape map and a region mask. The shape map specifies the desired volume and direction of the brush, while the region mask enables users to define specific areas to be generated.

Our experiments demonstrate that Text2VDM produces high-quality and diverse VDM brushes that can be directly integrated into mainstream modeling software, such as Blender~\cite{Blender:Community:1999} and ZBrush~\cite{ZBrsuh:Maxon:1999}. 
Compared to existing methods that directly generate full 3D models, our approach 
% of generating VDM brushes 
addresses a different use case where brush-based user sculpting is desirable. This enables artists to interactively use a variety of brushes to sculpt diverse and expressive models from a plain shape.

This paper makes the following contributions:
\begin{itemize}
    \item We first introduce the task of text-to-VDM brush generation, which is challenging to tackle directly using current text-to-image and text-to-3D methods.

    \item We propose Text2VDM, a novel framework for text-to-VDM brush generation that is readily compatible with artists' workflow of 3D asset creation.
    
    \item  We design a novel Semantic Enhancement SDS loss, which uses weighted blending to mitigate semantic coupling for sub-object structure generation.

% We introduce CFG-weighted blending in SDS, to model a more precise target distribution, in resolving the issue of semantic coupling in sub-object structure generation.
% We introduce CFG-weighted blending of tokens in the text prompt to SDS, which can effectively model a more precise target distribution, which resolve the issue of semantic coupling in sub-object structure generation.
% We propose to model a more precise target distribution in SDS via CFG-weighted blending to effectively resolve the issue of sematic coupling in sub-object structure generation.
% We leverage CFG-weighted blending for the tokens in the prompt, which can effectively resolve the issue of semantic coupling in SDS in sub-object structure generation.
\end{itemize}

%-------------------------------------------------------------------------

\begin{figure*}[!htb]
\centering
\includegraphics[width=0.96\textwidth]{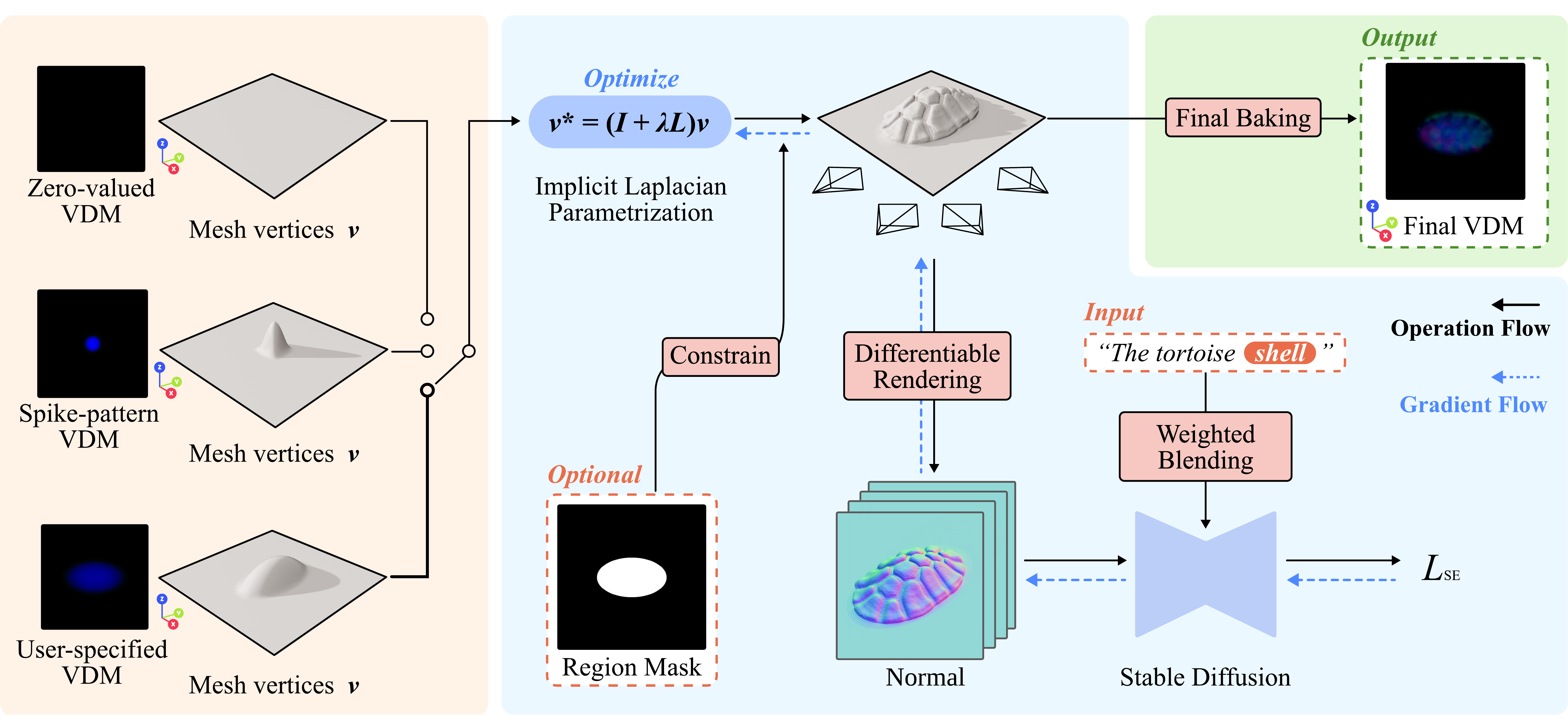}
\caption{
\textbf{Overview of Text2VDM.}
Starting with a dense planar mesh constructed from a zero-valued VDM, users can initialize the mesh volume through a default spike-pattern VDM or a user-specified VDM. Given the text prompt and region mask, we propose the Semantic Enhancement SDS loss $\mathcal{L}_{\text{SE}}$ to guide mesh deformations through a mesh Laplacian $L$ iteratively, achieving semantically focused generation of surface details or geometric structures.
% Laplace-Beltrami operator 
After generation, vertex displacements are baked into the final VDM.
}
\label{Fig: Pipeline}
\end{figure*}

\section{Related Work}
\label{sec:Relatedwork}
% This section reviews previous work related to 3D sculpting brush generation and summarizes the current research gap.

% Zeyu comment 1107: RW中每一个subsection，要先介绍子领域的发展概况，然后介绍几项关键工作，最后最重要的是，引出当前解决本论文中问题的Research gap（即现有方法不能解决我们的问题），突出我们问题和方法的创新性。
% \subsection{Text to Local 3D Generation and Editing}
\noindent\textbf{Text to Local 3D Generation and Editing.} With recent advances in diffusion models~\cite{StableDiffusion:Arxiv:2021} and differentiable 3D representations~\cite{DMTet:NIPS:2021,TriplaneDiffusion:CVPR:2023,SIGGRAPH:NJF:2022,NeRF:ECCV:2020,Largesteps:SIGGRAPH:2021}, many methods for text-guided full 3D model generation have emerged~\cite{Fantasia3D:ICCV:2023,SIGGRAPH:TextDeformer:2023,single1-instant3d,single7-richdreamer,single8-luciddreamer,CLIP-Forge:CVPR:2022,single12-text2mesh}. Since 3D content creation is an iterative process that often requires user interaction, more attention has been directed toward localized 3D generation and editing. For example, 3D Highlighter~\cite{3Dhighlighter} and 3D Paintbrush~\cite{paintbrush} use text as input, leveraging pre-trained CLIP models~\cite{CLIP:ICML:2021} or diffusion models~\cite{DreamFusion:ICLR:2022} to supervise the optimization of neural networks for segmenting the regions of a 3D model that match the text description. Based on the information from these segmented regions, further editing of texture and geometry can be applied to the 3D model. Furthermore, SKED~\cite{mikaeili2023sked} and SketchDream~\cite{SketchDream2024} introduce sketches as an additional modality to assist in localized editing. To enable more precise control, FocalDreamer~\cite{focaldreamer}, MagicClay~\cite{magiclay}, and Tip-Editor~\cite{tipeditor} allow users to specify the editing location directly within the 3D space. These works rely on optimization-based methods to edit specific objects, often resulting in non-reusable editing outcomes. Additionally, each edit requires a lengthy optimization process, making interactivity difficult to achieve. 

% To address it, we focus on generating reusable local geometric details and 3D component predefined sculpting brushes. These can be directly integrated into existing modeling workflows, allowing users to quickly perform localized edits to create exquisite 3D models.
% \vspace{-0.3 cm}
% \subsection{Diffusion Priors for 3D Generation}
% focus on the SDS improvements
\label{sec:related_sds}
\noindent\textbf{Diffusion Priors for 3D Generation.} Score distillation sampling (SDS)~\cite{SJC:CVPR:2023, DreamFusion:ICLR:2022} provides pixel-level guidance by seeking specific modes in a diffusion model, inspiring further research to improve optimization-based 3D generation~\cite{VSD, Magic3D:CVPR:2023, Perp-Neg:Arxiv:2023, ESD:Arxiv:2023, LODS:Arxiv:2023}. Some studies focus on mitigating the ``Janus'' problem~\cite{LMC-SDS:Arxiv:2024, Debias:NIPS:2023}, while others fine-tune diffusion models with multiview datasets to enhance 3D consistency~\cite{MVDream:Arxiv:2023, Zero123:ICCV:2023}. Recent research focuses on refining the design of SDS loss to achieve more precise guidance. For instance, Make-it-3D~\cite{Make3D:2023:ICCV} introduces two-stage optimizations to improve textured appearance, while Fantasia3D~\cite{Fantasia3D:ICCV:2023} dynamically modifies the time-dependent weighting function within SDS computations. Additionally, several methods~\cite{CSD:Arxiv:2023,NFSD:Arxiv:2023} incorporate negative prompts as the conditional term to further refine the optimizations. Although diffusion priors have achieved promising results, their application in generating sub-object structures without global context as a reference is still challenging.

% \subsection{Appearance and Geometric Brush Synthesis}
\noindent\textbf{Appearance and Geometric Brush Synthesis.} The concept of brushes is very common in the creative process of digital artists, serving as a reusable local decorative unit. Appearance brushes focus on color representation and drawing styles in 2D space. With the development of generative models~\cite{GAN:NIPS:2014,StableDiffusion:Arxiv:2021}, many works have explored the synthesis of procedural material~\cite{li2023end,li2024procedural} for 3D object texturing and appearance brushes for interactive painting~\cite{DiffusionTexturePainting,NeuralBrushstrokeEngine}, realistic artworks generation~\cite{stylizedneuralpainting,painttransformer,GeneralVS}, and applying stylization~\cite{RethinkingST,snps}. Unlike appearance brushes, geometric brushes focus on modifying geometry by moving the vertices of a mesh in 3D space. VDM brushes, as an extension of standard geometric brushes, provide more complex geometric effects by utilizing VDMs. To the best of our knowledge, only a few techniques adopted the concepts of VDM for generation~\cite{64x64pixels,headcraft} and geometric texture transfer~\cite{deepgeometrysys}. 
Recently, concurrent work~\cite{genvdm} explored using a single image as input and leveraged a diffusion model for generating multiview normal maps to guide VDM reconstruction. In comparison, our method is fundamentally different in tackling the semantic coupling problem arising from text guidance. This demonstrates that generating geometric brushes is a highly promising research direction.

\vspace{-0.2 cm}
\section{Methodology}
\label{sec:Method}
To generate VDM brushes compatible with mainstream modeling software, we begin by constructing a dense mesh from an initial VDM, as shown in \Cref{Fig: Pipeline}.
% and enable control over the mesh volume and the deformable region with a shape map \( S \) and a region mask \( R \) respectively.
% and enable control over the mesh volume with an optional user specified initial VDM shape map \( S \).
% % 对spike pattern的描述
% This effectively adjust the Laplacian term in ~\Cref{ls}, pivoting/steering the gradient direction of the mesh deformation for large 3D components.
% , while ensuring the generated mesh can be directly baked as a brush. 
% We reparameterize the mesh vertices through an implicit formulation with a Laplace-Beltrami operator~\cite{Largesteps:SIGGRAPH:2021} to ensure high-quality mesh deformation.
We then apply score distillation with Stable Diffusion to guide high-quality mesh deformation formulated as a form of Sobolev preconditioned optimization~\cite{Largesteps:SIGGRAPH:2021}.
% The deformable region of the mesh can be controlled by an optional user-specified region mask $R$.
% However, directly using the original score-distillation sampling (SDS) can result in semantic coupling when generating 3D components, leading to unintended extra parts connected to the text-described object. 
To produce the intended sub-object level structure described in the text, we design a Semantic Enhancement SDS loss by applying weighted blending to the tokens in the prompt, effectively handling the issue of semantic coupling in SDS.

\subsection{Brush Initialization}
% 初始化部分只说 shape map；在mesh deformation部分再提region mask
% 预设了一些模板,也提供了一个简单的交互方式控制给用户控制mesh的生长方向，背后的原理：更好的引导梯度下降的方向

We provide three methods to initialize a base mesh for brush generation via a zero-valued VDM, a spike-pattern VDM, or a user-specified VDM.
A VDM is represented as a $512\times512$ three-channel image, in which each channel stores the displacement in the X, Y, or Z direction, respectively.
We first construct a planar grid mesh by creating two triangles for every $2\times2$ pixels and then apply the displacement stored in the VDM to mesh vertices.
The values in these three initial VDMs range from 0 to 1, in which 0 represents no displacement, and 1 corresponds to half of the mesh's edge length in the positive axis direction.
Since users can apply sculpting brushes symmetrically, our initial VDM does not need to store any negative values.

Our three methods for brush initialization facilitate the generation of diverse sculpting brush styles.
The zero-valued VDM results in a planar mesh, which is our default setup when no control is provided.
The spike-pattern VDM is suitable for generating protruding geometric structures, as it can effectively adjust the Laplacian term in \Cref{ls} to steer the gradient direction for mesh deformation.
For better control of the brush's volume and direction, we also provide an interface for users to create custom VDMs, so the user-specified brush initialization can effectively guide mesh deformation toward the target structure.

% We provide three methods to initialize a base mesh for brush generation via a zero-valued VDM, a spike-pattern VDM, or a user-specified VDM.
% A VDM is represented as a $512\times512$ three-channel image, in which the values in each channel store the displacement in XYZ directions.
% We convert the VDM to a dense mesh by creating two triangles from every $2\times2$ pixels, i.e., without loss of generality, one triangle on the top-left and the other on the bottom-right.

% The first approach uses a zero-valued VDM to generate a dense planar mesh. The second approach offers a spike-patterned VDM, which effectively adjusts the Laplacian term in ~\Cref{ls}, steering the gradient direction of the mesh deformation for generating larger 3D components. For better control of the brush's volume and direction, we offer a simple interactive interface for creating custom VDMs, allowing users to create an initial volume and shaping the mesh toward the target sub-object level structure.
% To enable control over the mesh volume with an optional user specified initial VDM shape map \( S \).
% 对spike pattern的描述
% This effectively adjust the Laplacian term in ~\Cref{ls}, pivoting/steering the gradient direction of the mesh deformation for large 3D components.
% For better control of the brush's volume and direction,
% we allow users to provide control through a customize shape map that serve as a template VDM. This deformation creates an initial volume, shaping the mesh toward the target sub-object level structure. 

%%%%%%%%%%%%%%%%%%%%%%%%%%%%%%%%%%%%%%%%%%%%%%%%%%

\subsection{Brush Generation via Mesh Deformation}
\label{sec: mesh_deform}
%直接用SDS优化顶点位置这种方法们虽然能够获得较大的顶点移动，但是需要很强的几何平滑正则项来维持mesh的质量，这导致了优化过程极其不稳定，难以收敛，最后产生过噪或是过平滑的结果
Given the vertices $v$ on the initialized base mesh, our method aims to learn a mesh deformation to the target brush shape.
The vertex positions $\hat{v}$ after mesh deformation can be expressed by:
\begin{equation}
\label{eq1:vertex}
    \hat{v} = \argmin_{v} \mathcal{L}_{\text{SE}}(\mathcal{D}_c(v), y),
\end{equation}
where $c$ represents the camera setup in a differentiable renderer $\mathcal{D}$~\cite{Laine2020diffrast}.
The loss function $\mathcal{L}_{\text{SE}}$ receives the rendered normal image $\mathcal{D}_c(v)$ and text input $y$ to evaluate the semantic guidance, which is detailed in~\Cref{sec: tesds}.
To accompany the external forces from $\mathcal{L}_{\text{SE}}$ that drive the mesh deformation with intrinsic smoothness energies, we follow the framework of a Sobolev preconditioned gradient descent~\cite{Largesteps:SIGGRAPH:2021}, where the base mesh is reparameterized by the mesh Laplacian $L$:
\begin{equation}
\label{ls}
    v^{*} = (I+\lambda L)v.
\end{equation}
This preconditioning involves solving a sparse linear system at every iteration, 
modifying the gradient descent update for each mesh deformation step to:
\begin{equation}
v \leftarrow 
v -\eta (I +\lambda L)^{-1}
\frac{\partial{\mathcal{L}_{\text{SE}}}}{\partial{v}},
\end{equation}
where $\eta$ is the learning rate, $I$ is the identity matrix, and $\lambda$ is a hyperparameter to control the extent of gradient diffusion over the
entire domain.
% When $\lambda=0$, this representation degrades to direct vertex displacements. As $\lambda$ increases, the mesh deforms toward more global structural changes. 
We set $\lambda=15$ throughout our experiments to balance the global structure and fine details during mesh deformation. 

\begin{figure}[tbh]
\centering
\includegraphics[width=0.48\textwidth]{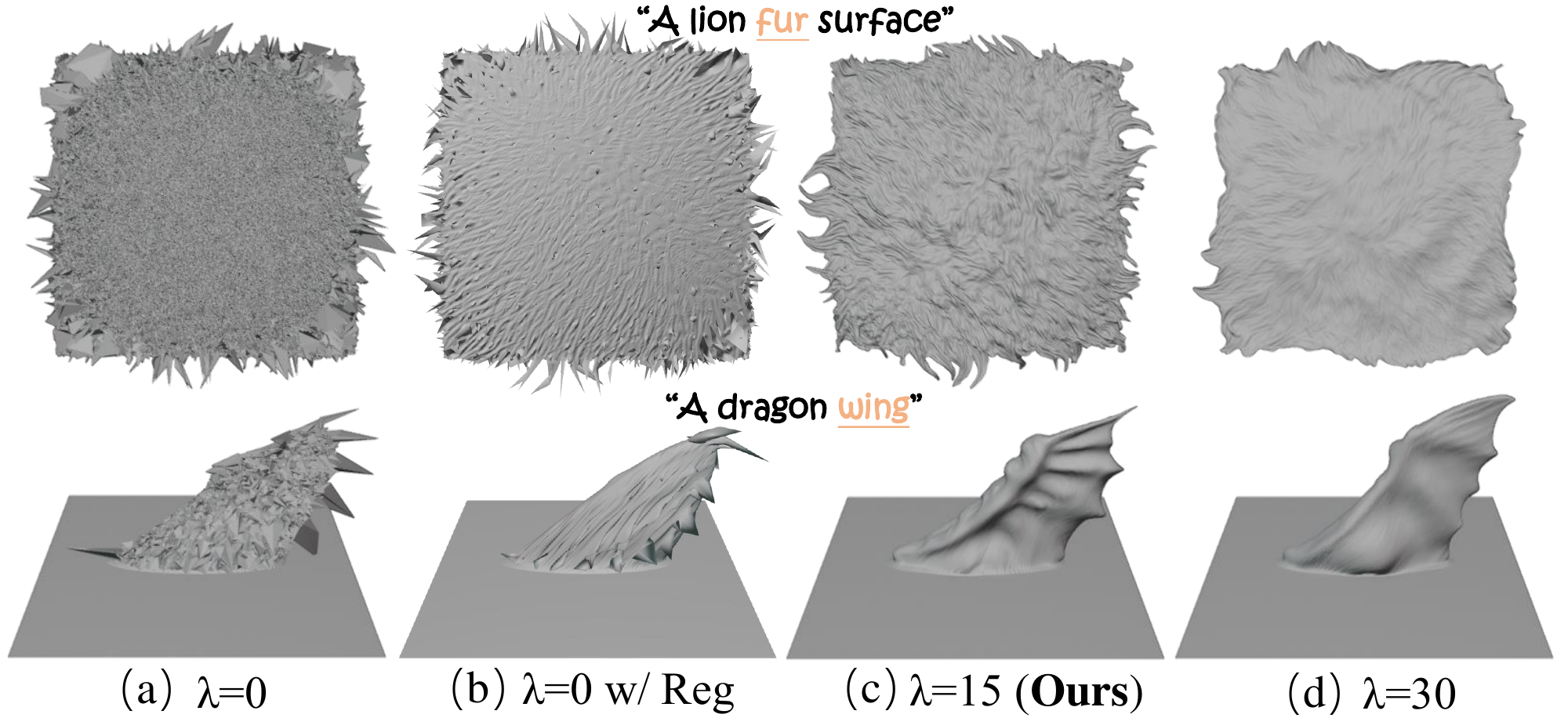}
\caption{
    \textbf{Effect of $\lambda$.} (a) Low-quality results w/o mesh smoothness, (b) low-quality results w/ direct Laplacian regularization, (c) our adopted preconditioning scheme with $\lambda$ = 15, and (d) results with over-smoothness caused by a larger $\lambda$.
}
\label{Fig: lambda}
\end{figure}

Compared to directly adding a Laplacian regularization term to \Cref{eq1:vertex}, the preconditioning framework~\cite{Largesteps:SIGGRAPH:2021} is critical in achieving large mesh deformation while maintaining proper topology with reduced triangle flips (\Cref{Fig: lambda}). 
%Please see \Cref{Fig: lambda} for the comparison as well as the choices of $\lambda$.
% In mesh deformation strategies, directly applying displacement to each vertex often results in unintended self-intersections of mesh faces caused by noisy gradients from pixel-level losses~\cite{single12-text2mesh}. 
% To address it, s
Several works~\cite{SIGGRAPH:TextDeformer:2023,HeadEvolver:Arxiv:2024} adopt the strategy by Aigerman et al.~\cite{SIGGRAPH:NJF:2022}, parameterizing deformation through Jacobian fields that capture the scaling and rotation of each triangle. 
Although this method effectively smooths vertex displacements, the local deformation represented in Jacobians accumulates, leading to global drifting for open-boundary meshes, making it challenging to bake the mesh as a brush.
Additionally, we provide an optional region mask to restrict mesh deformation to the user-defined region during optimization. It helps maintain zero values in unused VDM areas when generating geometric structures. For producing surface details, the region mask ensures that the brush effect meets user-customized requirements (\Cref{Fig: Effect of region mask}).
% By adjusting the activation ratio of the region mask, the final brush effect can effectively match the user's guidance. For instance, our experiment activated the region mask for the first half of total iterations as a warm-up stage to effectively control the shape of the surface detail (\Cref{Fig: Effect of region mask}). 
% and allowing the overall effect of the brush to align with the user's guidance (\Cref{Fig: Effect of region mask}). 
% This differentiable parameterization effectively alters the gradient propagation at each optimization step as:
% \begin{equation}
% v^{*} \leftarrow v^{*} -\eta (I +\lambda L)^{-1}\frac{\partial{\mathcal{L}_{\text{WS}}}}{\partial{v}^{*}}
% \end{equation} 
%\begin{equation}
%    v^{*} \leftarrow v-\eta (\frac{\partial{\mathcal{L}_{\text{WS}}}}{\partial{v}}+\lambda Lv),
%\end{equation}
% where $\eta$ is the learning rate.
% $\tau$ represents the current iteration step, and $\mu$ is the ratio of activated region mask iterations within the total iterations $\kappa$. In our experiments, $\mu$ is set to 1 for generating 3D components and to $\frac{1}{2}$ for generating surface styles. 

By framing the VDM generation task as mesh deformation through preconditioned optimization, our framework preserves the favorable initial mesh topology and ensures control throughout the procedure.
% The advantage of this Laplacian energy-aware mesh deformation is that it enhances the robustness of optimization for non-convex objective functions. 
The resulting mesh preserves the original structure while incorporating rich local deformations, making it well-suited for baking as a brush. 
% Furthermore, this parameterization is not constrained by shape priors or even mesh representations, making it highly versatile. It has been successfully validated to support texture reconstruction~\cite{Largesteps:SIGGRAPH:2021}, which aligns well with our need for VDM generation. Based on the experimental conclusions of Haetinger et al.~\cite{MeshTransfer:SIGGRAPH:2024}, $\lambda$ was set to 20, enabling mesh deformation for both large structures and local details. 

% \vspace{-0.6 cm}
\subsection{Semantic Enhancement Score Distillation}
\label{sec: tesds}
% 先介绍原始的sds，公式（类似preliminary）（参考ThemeStation）
% sds在生成3Dcomponent出现的问题（语义耦合），指出现象，分析一下原因
Current text-to-3D generation methods like DreamFusion~\cite{DreamFusion:ICLR:2022} often optimize a 3D representation parameterized by \( \theta \) so that rendered images \( \mathbf{x} = g(\theta) \) resemble 2D samples produced by a pre-trained T2I diffusion model for a given text prompt \( y \). \( g \) functions as a differentiable renderer. The T2I diffusion model \( \phi \) predicts the sampled noise \( \epsilon_{\phi}(\mathbf{x}_t; y, t) \) of a rendered image \( \mathbf{x}_t \) at a noise level \( t \) for the text input \( y \).
To make rendered images follow the text-conditioned distribution in Stable Diffusion, the SDS loss updates \( \theta \) by estimating the gradient:
\begin{equation}
    \nabla_{\theta} \mathcal{L}_{\text{SDS}}(\phi, \mathbf{x}) = \mathbb{E}_{t, \epsilon,c} \left[ \omega(t) \left( \epsilon_{\phi}(\mathbf{x}_t; y, t) - \epsilon \right) \frac{\partial \mathbf{x}}{\partial \theta} \right],
\end{equation}
where $\omega(t)$ is a time-dependent weighting function.

However, the SDS loss cannot effectively supervise sub-object structure generation due to the issue of semantic coupling in full objects.
For example, when using the SDS loss to generate a tortoise shell, it also usually generates the tortoise's tail and head, causing semantic coupling (\Cref{Fig: Effect of CFG-weighted SDS}).
We believe that the issue of semantic coupling in SDS stems from the training data of Stable Diffusion, in which most images contain full objects rather than separate parts.
Therefore, the semantics of full objects often appear in the target distribution conditioned on text only describing sub-object structures.
% of sub-object structures containing semantic information related to the full object.
% SDS gradients containing guidance information associated with semantics beyond those described by the input text. 

A straightforward approach is using negative distributions via Classifier Score Distillation (CSD)~\cite{CSD:Arxiv:2023} or Variational Score Distillation (VSD)~\cite{VSD} to suppress coupled semantics. CSD employs predefined negative prompts, resulting in more accurate negative distributions than those adaptively learned by VSD~\cite{CSD:Arxiv:2023}:
% demonstrates that compared to Variational Score distillation (VSD)~\cite{VSD}, which adaptively learns negative classifier scores, CSD employs predefined negative prompts, resulting in a more precise optimization process:
%\begin{equation}
%\footnotesize
%    \delta^{cls}_x = \omega_{1} \cdot \epsilon_{\phi}(\mathbf{x}_t; y, t) + (\omega_{2} - \omega_{1}) \cdot \epsilon_{\phi}(\mathbf{x}_t; t) - \omega_{2} \cdot \epsilon_{\phi}(\mathbf{x}_t; y_{neg}, t),
%\end{equation}
\begin{align} % the 
\label{csd}
\nabla_{\theta} \mathcal{L}_{\text{CSD}}(\phi, \mathbf{x}) &= \mathbb{E}_{t, \epsilon,c} [(\omega_{\text{pos}} \cdot \epsilon_{\phi}(\mathbf{x}_t; y, t) \nonumber \\
&- \omega_{\text{neg}} \cdot \epsilon_{\phi}(\mathbf{x}_t; y_{\text{neg}}, t))\frac{\partial \mathbf{x}}{\partial \theta}],
\end{align}
where $\omega_{\text{pos}}$ and $\omega_{\text{neg}}$ denote different weights for positive and negative prompts.
It requires two separate inferences with positive and negative prompts to obtain two distributions, which are then subtracted to suppress coupled semantics. However, our experiments show that CSD is ineffective in decoupling semantics because the negative prompt cannot accurately represent the undesirable coupled semantics in the positive prompt (\Cref{Fig: Effect of CFG-weighted SDS}). This results in noisy guidance, making CSD less effective in decoupling semantics.
% A similar issue was also mentioned by Hertz et el.~\cite{p2p}. 
% We found that the negative prompt's semantic distribution does not align with the associated semantics in the target distribution conditioned by the positive prompt of sub-object structures.
% We found that the negative semantic distribution conditioned by the negative prompt does not align with the associated semantic distribution of full object in the target distribution conditioned by the positive prompt describing sub-object structure.
% During our experiments, we observed that the negative prompt added a term that lies outside the distribution domain of the positive prompt. 
% This resulted in unstable target distribution, making CSD less effective at decoupling semantics. 
% Furthermore, as the weight of the negative prompt increased, the optimization became more unstable and challenging to converge.

Unlike semantic suppression in CSD, we propose a semantic enhancement method to mitigate semantic coupling by enhancing the semantics of part-related words.
This can lead to a more accurate and stable target distribution, as shown in \Cref{Ablation}. Our key design is to apply weighted blending to the tokens in the original prompt to obtain a semantically focused text embedding, which serves as stable guidance for the optimization process.
% Unlike CSD, we apply weighted blending to the tokens in the original prompt, which does not require additional inference to construct a negative distribution. This results in semantically focused text embedding, directing toward a more precise target distribution.
% It enhances the semantics of local components within the original prompt to achieve a more precise distribution. 
We define the Semantic Enhancement SDS loss as:
\begin{equation}
\small
    \nabla_{\theta} \mathcal{L}_{\text{SE}}(\phi, \mathbf{x}) = \mathbb{E}_{t, \epsilon,c} \left[ \omega(t) \left( \epsilon^{*}_{\phi}(\mathbf{x}_t; y, t) - \epsilon \right) \frac{\partial \mathbf{x}}{\partial \theta} \right],
\end{equation}
%\begin{equation}
%\small
%    \mathcal{L}_{\text{WS}}(\phi, \mathbf{x}) = \mathbb{E}_{t, \epsilon} \left[ \omega(t) \left( \epsilon_{\phi}(\mathbf{x}_t; y_{\text{CFG}}, t) - \epsilon \right)  \right],
%\end{equation}
where $\epsilon^{*}_{\phi}(\cdot)$ uses a text embedding weighted by Compel~\cite{Compel:github:2023}. 
Specifically, we assign a weight $s$ to each word in the prompt and compute the weighted embedding $e_\text{w}$ for each word by blending the original word embedding $e$ and the empty text embedding $e_{0}$: $e_{\text{w}} = e_{0} + s \cdot (e - e_{0})$. By concatenating the weighted embedding of each word in sequence, we obtain the final semantically focused text embedding.
% Specifically, we assign each word in the prompt a weight $s$ and compute the weighted embedding $e_w$ for each word by blending it with the empty text embedding $e_{\phi}$ as follows: $e_w = e_{\phi} + s\cdot 
% (e - e_{\phi}) $. By concatenating the weighted embeddings of each word in sequence, we obtain the final semantically focused text embedding $y^{*}$. $1.1^2$
In our experiments, we found that using 
% $s=1.21$ 
$s={1.1}^{2}$ for words that require semantic enhancement can achieve stable optimization and effectively alleviate the issue of semantic coupling.

\begin{figure}[tbh]
\centering
\includegraphics[width=0.46\textwidth]{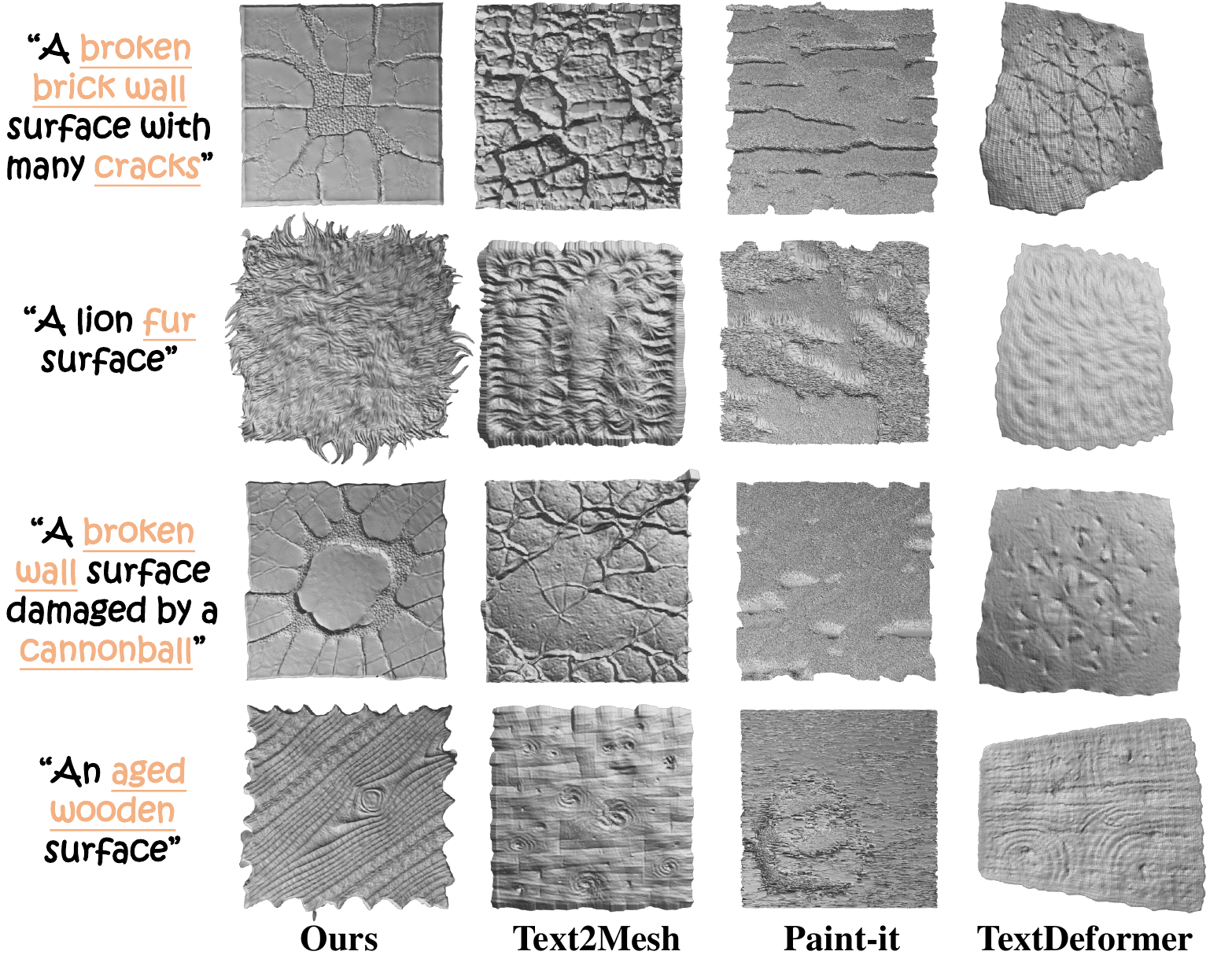}
\caption{
    \textbf{Qualitative comparisons of generated brushes for surface details.} Our method captures geometry details guided by text, effectively preserving the surface structure and avoiding mesh distortion.
}
\label{Fig: Qualitative 2D}
\end{figure}
% \vspace{-0.6 cm}
\section{Experiments}
\label{sec:Experiment}
We conducted experiments to evaluate the various capabilities of Text2VDM both quantitatively and qualitatively for text-to-VDM brush generation.
% in ~\Cref{Qualitative} and ~\Cref{Quantitative}.
We then present an ablation study that validates the significance of our key insight into Semantic Enhancement SDS, as well as the effect of the region mask and VDM initialization.
% in ~\Cref{Ablation}.

\begin{figure*}[tbh]
\centering
\includegraphics[width=0.96\textwidth]{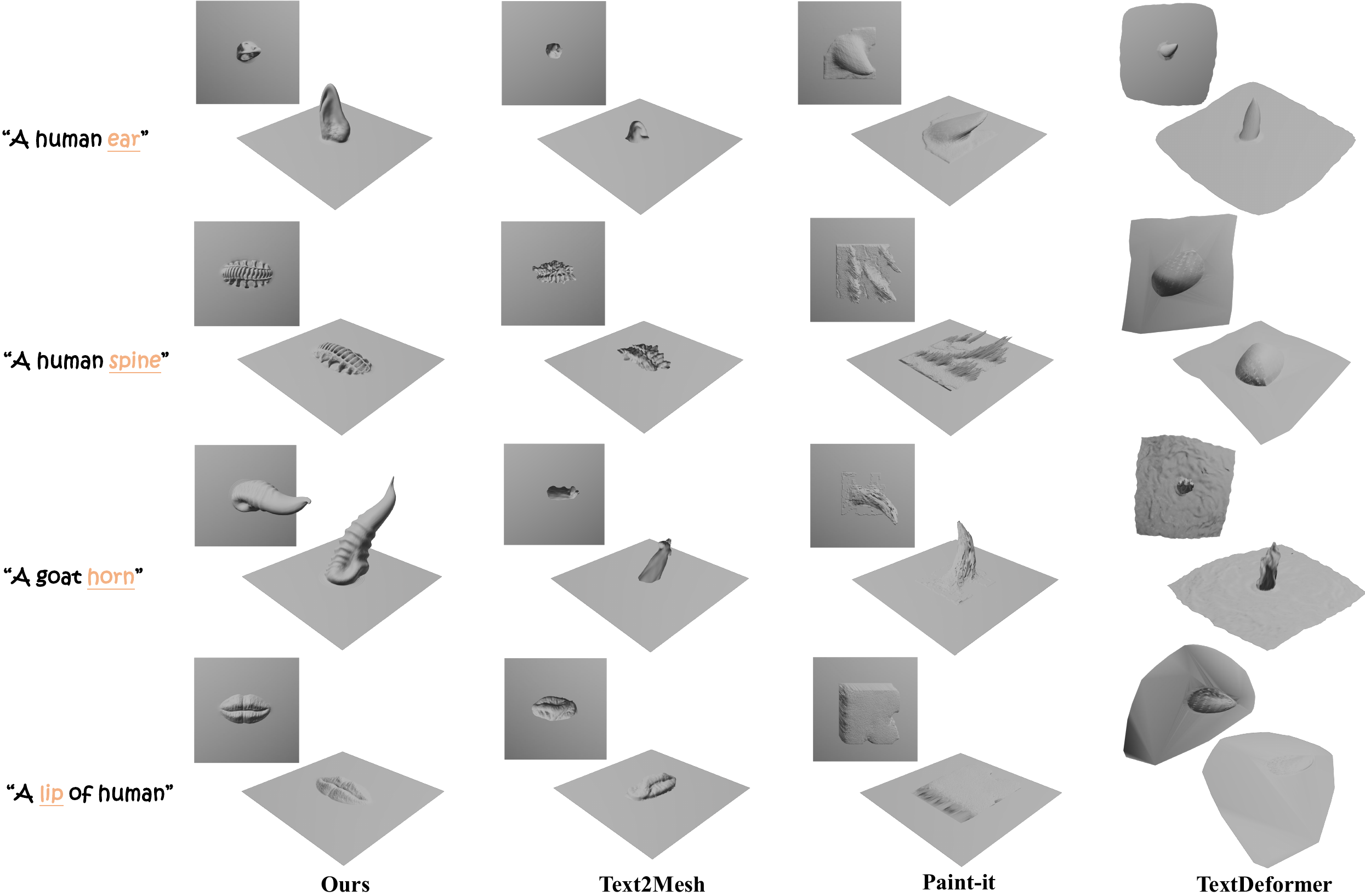}
\caption{
    \textbf{Qualitative comparisons of generated brushes for geometric structures.} Our method accurately presents key geometric features described by text, facilitating downstream applications in modeling software.
}
\label{Fig: Qualitative 3D}
\end{figure*}

\subsection{Qualitative Evaluation}
\label{Qualitative}
To the best of our knowledge, Text2VDM is the first framework to generate VDM brushes from text.
We adapted three existing methods for comparison and classified them into two categories. The first category includes Text2Mesh~\cite{single12-text2mesh} and TextDeformer~\cite{SIGGRAPH:TextDeformer:2023}, which generate a brush mesh through text-guided mesh deformation on a planar mesh, following a process similar to ours. For the second category, we opt to directly generate VDM via Paint-it~\cite{paintit}. Notably, this method originally uses SDS to optimize a UNet for generating PBR textures. We reframed it to suit our VDM brush generation task, modifying it to generate VDM through SDS optimization of the UNet. In geometric structures generation experiment (\Cref{Fig: Qualitative 3D}), all methods are compared fairly, with the same non-zero VDM initialization and mask applied to each prompt (see Appendix B.2). For surface details (\Cref{Fig: Qualitative 2D}), all methods start with a zero-valued VDM and no masks to ensure a fair comparison. 
% We compared the visual results in \Cref{Fig: Qualitative 2D} and \Cref{Fig: Qualitative 3D}.
% In \Cref{Fig: Qualitative 3D}, to ensure a fair comparison, we used the same VDM initialization. See Appendix B.2 for details.

Compared to other methods, Text2VDM can generate better-quality VDM brushes. Text2Mesh applies displacement to each vertex along normal directions, resulting in limited mesh deformation. TextDeformer indicates the accumulation of local deformations in the Jacobians, which results in global mesh drift, making it challenging to bake these meshes into VDM.
Reframed Paint-it VDM generation is equivalent to optimizing the three-axis displacement of each vertex on the mesh with SDS. Although the UNet reduces noise from the SDS~\cite{paintit}, smooth regularization is still required to ensure mesh quality, which makes achieving high-quality mesh generation quite challenging.
% Using a UNet to generate VDM is equivalent to optimizing the three-axis displacement of each vertex on the mesh with SDS. Although the UNet reduces noise from the SDS~\cite{paintit}, geometric regularization is still required to ensure mesh quality. The generated mesh must compromise between solving the problem and being smooth, which results in low-quality mesh generation.

\subsection{Quantitative Evaluation}
\label{Sec: Quantitative}

We quantitatively evaluated our framework regarding generation consistency with text input and mesh quality. We used 40 distinctive text prompts for VDM generation.

\noindent\textbf{Generation Consistency with Text.} We initially assessed the relevance of the generated results to the text descriptions~\cite{CLIP:ICML:2021}. 12 different views were rendered for average scores respectively, as presented in Table~\ref{tab:quantitative comp}. Our approach achieves the highest scores compared to baseline methods.

\begin{table}[h!]
\caption{Quantitative evaluation of state-of-the-art methods. The geometry CLIP score is calculated on shaded images with uniform albedo colors~\cite{Richdreamer:CVPR:2024}, and self-intersection is quantified as the ratio of self-intersected mesh faces to the total number of faces.}
\centering
\footnotesize   % incase not overflow
% TADA & TextDeformer & Fantasia3D
\begin{tabular}{*{10}{c}}
         \hline
           & Geometry CLIP Score $\uparrow$ & Mesh Self-Intersection $\downarrow$\\
         \hline 
         Paint-it & $0.2375$ & $19.42\%$  \\
         Text2Mesh & \underline{0.2497}  & $7.18\%$\\
         TextDeformer & $0.2477$  & \textbf{0.04\%} \\
         Ours & \textbf{0.2556} &  \underline{0.77\%}\\
         \hline
\end{tabular}
\label{tab:quantitative comp}
\end{table}

\begin{figure*}[tbh!]
\centering
\includegraphics[width=0.95\textwidth]{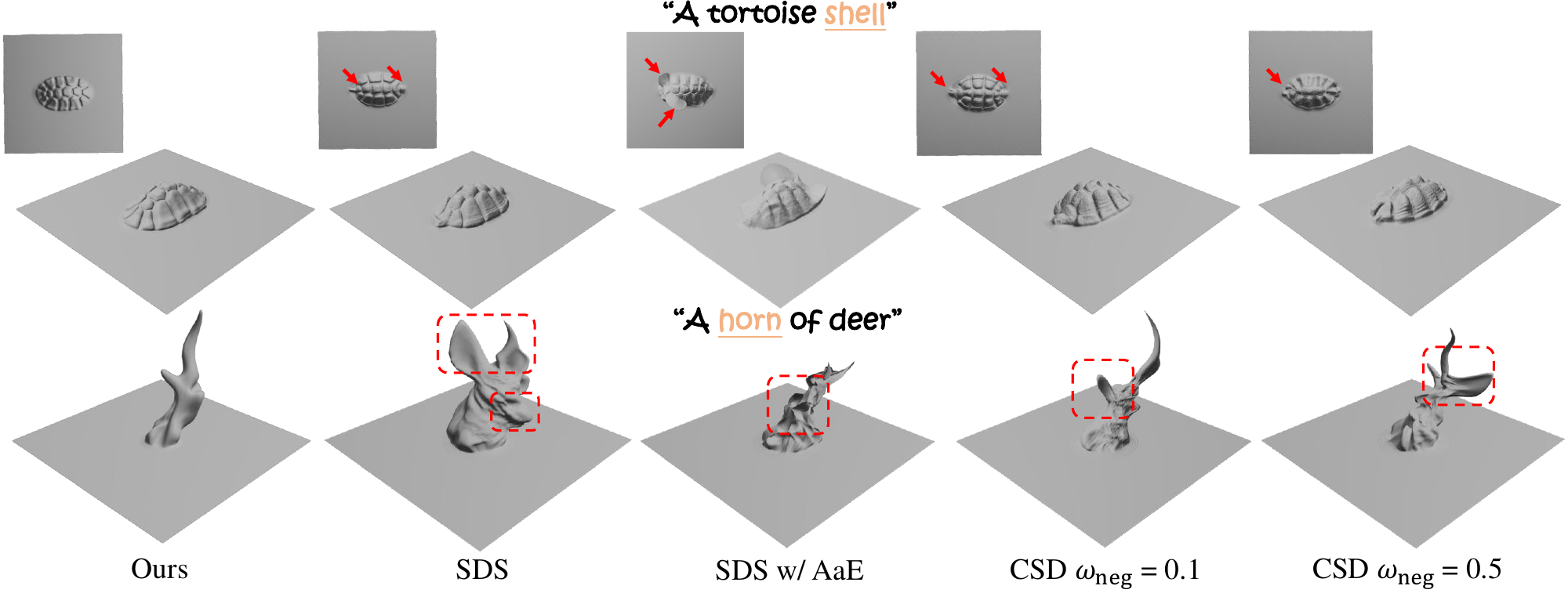}
\caption{
    \textbf{Effect of Semantic Enhancement SDS.} Our method effectively mitigates semantic coupling issues in SDS, such as generating the tortoise’s tail and head or the deer’s ear and mouth, by providing more focused semantic guidance. In contrast, the semantic suppression method used in CSD and the other semantic enhancement approach proposed by Attend-and-Excite both lead to an unstable optimization.
} 
\label{Fig: Effect of CFG-weighted SDS}
\end{figure*}

\noindent\textbf{Mesh Quality.} We evaluated mesh quality by examining self-intersection. Paint-it and Text2Mesh, which utilize direct vertex displacement, often converge to a local minimum and disregard the mesh triangulation. While TextDeformer exhibits the lowest self-intersection, its tendency to produce over-smoothed results frequently results in losing object features described in text prompts. 

\begin{table}[h!]
\caption{User evaluation of generated VDMs.}
\centering
\footnotesize
    \begin{tabular}[width=1.0\textwidth]{*{10}{c}}
         \hline 
         User Preference $\uparrow$  & Geometry Quality & Consistency  with Text\\
         \hline 
         Paint-it & $3.1\%$  & $1.7\%$ \\
         Text2Mesh & \underline{$18.3\%$} & \underline{$27.3\%$} \\
         TextDeformer & $3.3\%$ & $3.4\%$ \\
         Ours & \textbf{75.3\%} & \textbf{67.6\%} \\
         \hline
    \end{tabular}
\label{tab:user comp}
\end{table}

\noindent\textbf{User Study.} We further conducted a user study to evaluate the effectiveness and expressiveness of our method. A Google Form was utilized to assess 1) geometry quality and 2) consistency with text. We recruited 32 participants, of whom 14 are graduate students majoring in media arts, and 18 are company employees specializing in AI content generation. The participants were instructed to choose the preferred renderings of VDM from different methods in randomized order, as shown in Table~\ref{tab:user comp}. The results show participants preferred our method by a significant margin. 
% For practical evaluation, we invited 5 participants to use VDMs generated by our methods in Blender to sculpt 3D models that aligned with their expectations (Figures~\ref{Fig: Local to Global Mesh Stylization} and~\ref{Fig: Coarse to Fine Interactive Modeling}).
% `` ''
% \vspace{0.3cm}
\subsection{Ablation Study}
\label{Ablation}
\begin{figure}[tbh!]
\centering
\includegraphics[width=0.48\textwidth]{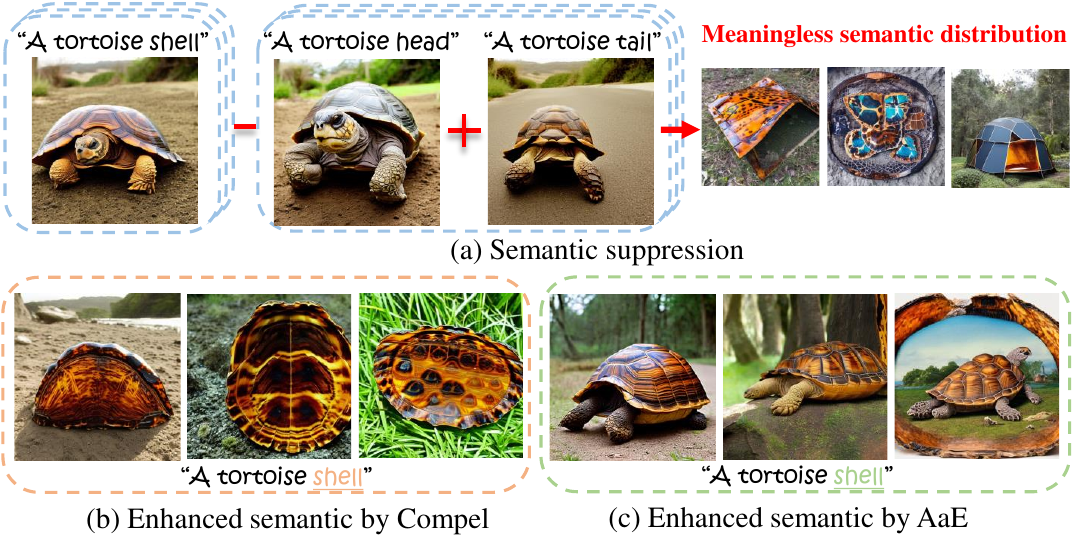}
% \vspace{0.1 cm}
\caption{
    \textbf{Visualization of semantic decoupling performance.} The images generated by the T2I model qualitatively visualize sub-object semantics. (a) Coupled semantics in positive and negative prompts lead to meaningless distributions when subtracted. (b) Enhanced semantics by Compel achieves stable decoupling. (c) AaE can only ensure that the enhanced semantics are preserved, but cannot decouple negative ones effectively.
    % 我们用stable diffusion模型生成的图片定性代表对应语义. (a) 由于positive prompt和negative prompt都会有耦合语义导致相减之后会得到无意义得语义分布. (b) 我们的方法可以得到稳定的解耦的语义. (c) Attend-and-Excite 只能保证被强调的语义不会丢失，但是无法很好的对语义进行解耦。
}
\label{Fig: analysis}
\end{figure}

\textbf{Effect of Semantic Enhancement SDS.} \Cref{Fig: Effect of CFG-weighted SDS} compares the results generated by the original SDS~\cite{DreamFusion:ICLR:2022}, our Semantic Enhancement SDS with Compel, SDS with another semantic enhancement method in Attend-and-Excite (AaE)~\cite{aae}, and CSD~\cite{CSD:Arxiv:2023} with two different annealed weights for the negative prompts: ``tortoise tail, tortoise head'' and ``deer's ear, deer's mouth.''
We also qualitatively compare these methods by visualizing their performance on semantic decoupling using the same Stable Diffusion for 2D image generation (\Cref{Fig: analysis}).
Compared to semantic suppression using negative prompts and semantic enhancement by AaE, our design of introducing a semantically focused text embedding via Compel to SDS is most effective in resolving the issue of semantic coupling.
% Through analysis, we find that using Compel for semantic enhancement is the most effective in mitigating semantic coupling in SDS.
% We use the same T2I model to qualitatively visualize semantic decoupling effect across above methods (\Cref{Fig: analysis}).

\noindent
\textbf{Key Insight.} As discussed in \Cref{sec: tesds}, SDS can result in semantic coupling when generating sub-object structures, leading to artifacts like the tortoise's tail and head or the deer's ear and deer's mouth.
We observed that the semantics represented by negative prompts are also coupled. Therefore, meaningless semantics can emerge when applying semantic suppression in SDS, which leads to unstable optimization. Increasing the weight of negative prompts further reduces the overall quality of generated results.
AaE enhances the cross-attention map of specific tokens by continuously updating the latent code.
However, AaE is not suitable for the SDS framework because the latent code is affected by different camera poses and random noise at each iteration.
% at each iteration comes from a randomly time-stepped noisy rendered image.
This temporal randomness undermines the continuity of the latent code, resulting in unstable optimization and subpar results.
In contrast, our method uses Compel to enhance semantics and achieves more effective semantic decoupling.
Moreover, the semantically focused text embedding produced by Compel is independent of temporal variation.
These properties help mitigate semantic coupling during SDS optimization, leading to high-quality sub-object geometric structures.
Meanwhile, our method can also enhance the semantics in the text prompt for surface detail generation (see Appendix A.2).
% while also ensuring that surface details better align with the enhanced semantics in the text prompt (see Appendix A.2).
% 我们发现负文本所代表的语义也有耦合情况，他不能准确的表示正提示中不想要的耦合语义，当两者相减时，会出现无意义的语义，导致优化过程不稳定，当增大负文本的权重时，会导致更低质量的结果
% Attend-and-Excite依赖持续更新的初始潜在向量（latent）来增强特定token的交叉注意力图（cross-attention map）。然而在SDS框架中，每次迭代输入的潜在向量均来自随机时间步加噪的渲染图像，这种时序随机性破坏了潜在空间的连续性.This results in unstable optimization, and subpar results.
% 我们使用的语义增强方法可以得到更加稳定的解耦后的语义，并且这种语义增强与时序无关，这使得我们的方法能够生成high-quality meshes
% Additionally, our method better aligns surface details with the text description (see Appendix A.2).
% Additionally, we find that when generating surface details, our method produces results that better align with the text description (see Appendix A.2). 
% We also found that using negative prompts was ineffective at decoupling semantics. Increasing the initial weight of negative prompts further makes the optimization unstable, resulting in low-quality results. 
% Due to the random time steps in each SDS iteration, the temporal randomness disrupts the continuity of the latent space, causing instability in the semantic enhancement of Attend-and-Excite. This results in unstable optimization, lack of detail, and mesh distortion.
% In contrast, our method effectively mitigates semantic coupling to produce high-quality meshes. % 另外我们也发现在生成surface detail时，使用我们的方法可以得到更符合文本描述的结果，please refer to Appendix for more details. 

% attend and excited 这个方法能够部分减缓语义耦合但是会导致不稳定优化 缺乏细节并伴随网格畸变，具体有两个原因，第一，原方法依赖持续更新的初始潜在向量（latent）来增强特定token的交叉注意力图（cross-attention map）。然而在SDS框架中，每次迭代输入的潜在向量均来自随机时间步加噪的渲染图像，这种时序随机性破坏了潜在空间的连续性。其次, 三维生成过程中，同一语义token在不同视角下对应的注意力图存在差异。这种空间不一致性导致语义增强效果随视角变动产生波动，难以维持稳定的特征强化。
\begin{figure}[tbh!]
\centering
\includegraphics[width=0.48\textwidth]{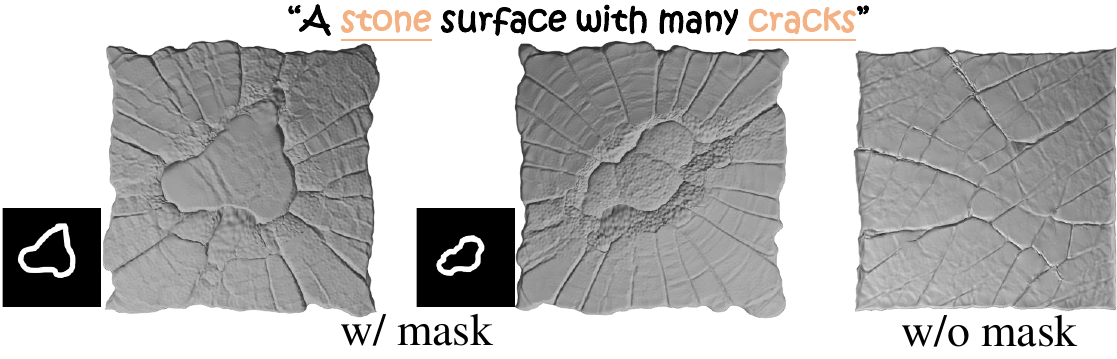}
\caption{
    \textbf{Effect of region mask.} Region masks can effectively control the pattern of surface details based on text input.
}
\label{Fig: Effect of region mask}
\end{figure}

% \vspace{-0.2 cm}
\noindent\textbf{Effect of Region Mask.}
\Cref{Fig: Effect of region mask} demonstrates two region masks and their control over surface detail generation given a text prompt. Without a region mask, the results can still match the text prompt but lack a specific pattern.
By using a region mask and maintaining an activation ratio of 1/2 throughout the total iterations as a warm-up stage, we achieve a reasonable tradeoff between mask-result alignment and generation diversity (see Appendix A.3). 
% The results align well with user guidance.

% By using a region mask and adjusting the activation ratio to 1/2 throughout the experiments of the region mask, we can strike a balance between alignment and diversity. please see Appendix A.3 for more details.  and the generated results effect can effectively match the user's guidance. 
% our generated results effectively conform to the user's desired patterns while also matching the styles, like metal and stone.

\begin{figure}[!htb]
\centering
\includegraphics[width=0.48\textwidth]{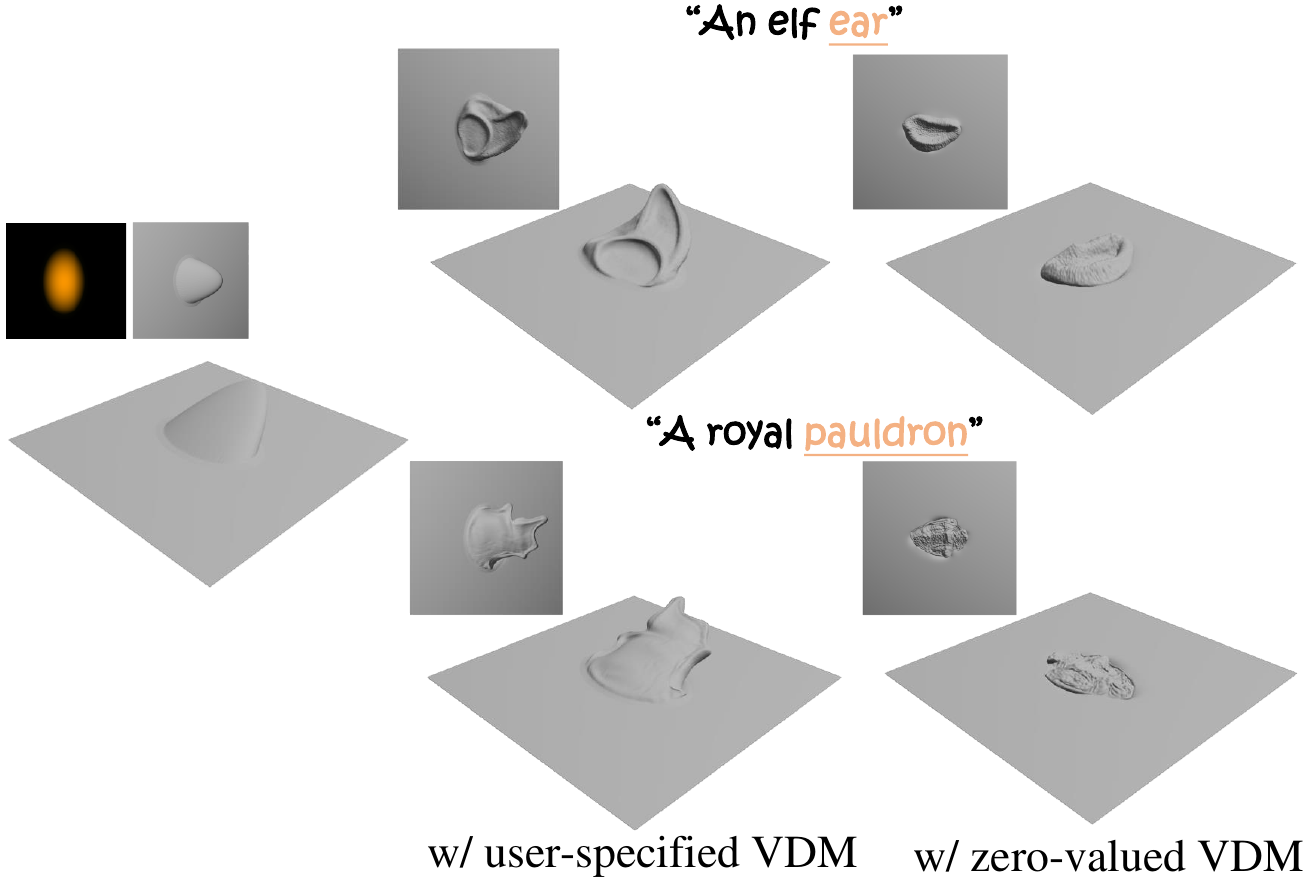}
\caption{
    \textbf{Effect of VDM initialization.} User-specified VDMs can help achieve the intended final effect of geometric structures by initializing the brush's volume and direction.
}
\label{Fig: Effect of shape map}
\end{figure}

\noindent\textbf{Effect of VDM Initialization.} % 我们的方法在图中展示了生成结果与通过shape map进行初始化体积与方向保持一致的能力，在不同的local component生成中，比如beard，pauldron，elf ear,我们的方法能够在保持体积和方向大致稳定的情况下生成多样的符合文本描述的结果
Our method demonstrates that user-specified VDMs can effectively control the volume and direction of generated geometric structures. \Cref{Fig: Effect of shape map} shows that the results are high-quality and match the text descriptions well, such as the elf ear and pauldron.
% We also found that without volume initialization, it is challenging to generate the desired results.
As this initializes the Laplacian term and steers the gradient flow in geometric structure generation, users can easily specify an initial VDM or choose a VDM template provided in our framework to generate expressive results.

\begin{figure}
\centering
\includegraphics[width=0.48\textwidth]{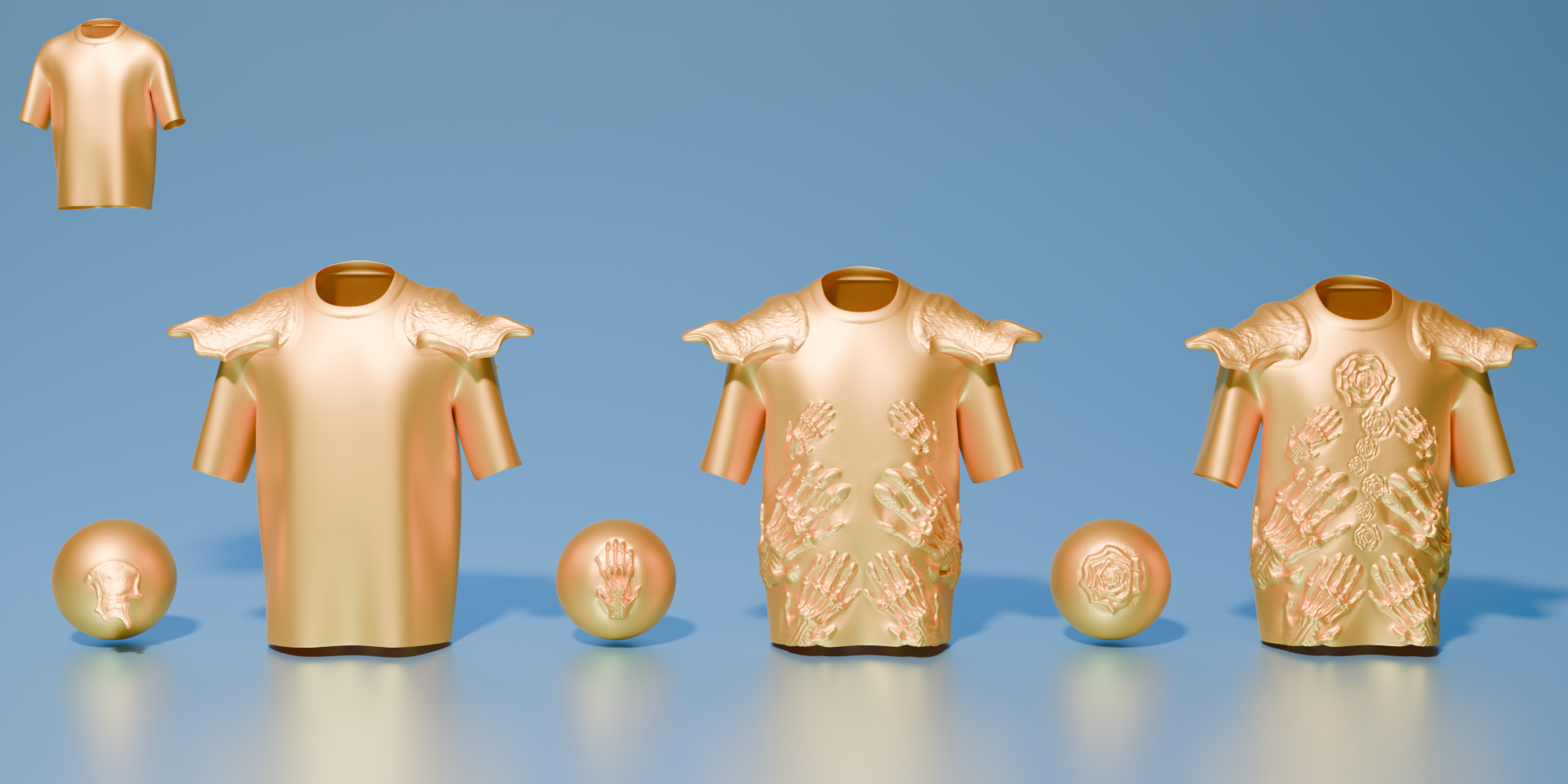}
\caption{\textbf{Coarse to fine interactive modeling.} By combining geometric structures brushes and surface details brushes for iterative sculpting in modeling software, users can rapidly create an expressive model from a plain shape (top left).}
\label{Fig: Coarse to Fine Interactive Modeling}
\end{figure}

% \vspace{-0.6 cm}
\subsection{Applications}
\label{Application}
Once various VDMs are generated, users can use these brushes to meet diverse creative needs in modeling software. For example, they can apply VDM brushes for mesh stylization and a real-time iterative modeling process.

\noindent\textbf{Local-to-Global Mesh Stylization.} Although mesh stylization is a complex task even for professional artists, combining different surface details allows users to achieve stylization quickly (see Appendix C.3). 
% For instance, users can apply a variety of wall-damage brushes to specific areas of a stone pillar, creating a style of damage (~\Cref{Fig: Local to Global Mesh Stylization}).
% Similarly, they can use different rust-effect brushes on a helmet to give it an aged style, 

% \begin{figure}[!htb]
% \centering
% \includegraphics[width=0.48\textwidth]{sec/Figures/application.pdf}
% \caption{\textbf{Local to global mesh stylization.} Applying various surface details brushes can create a damaged-style stone pillar model.} 
% \label{Fig: Local to Global Mesh Stylization}
% \end{figure}

\noindent\textbf{Coarse-to-Fine Interactive Modeling.} Unlike previous methods~\cite{magiclay,tipeditor} that require a lengthy optimization process for each edit and result in non-reusable outcomes, our generated VDM brushes can be directly used in modeling software. This enables users to apply the generated brushes easily and interactively (\Cref{Fig: Coarse to Fine Interactive Modeling}). 
% For example, \Cref{Fig: Coarse to Fine Interactive Modeling} shows that users can combine various brushes, such as skeleton hand, rose pattern, and pauldron to refine a coarse cloth model into a highly detailed one.

%\input{sec/5_Application}

\vspace{-0.1 cm}
\section{Conclusion}
\label{sec:Conclusion}
We have presented Text2VDM, a novel framework for VDM brush generation from text. A VDM is a non-natural 2D image where each pixel stores a 3D displacement vector, making it challenging for existing T2I models to generate. Therefore, we treat VDM generation as diffusion-guided mesh deformation formulated as a form of Sobolev preconditioned optimization.
% This allows users to engage in a real-time, iterative 3D model creation process. 
To mitigate semantic coupling issues in SDS, we propose using weighted blending for prompt tokens, achieving high-quality brush generation. Moreover,  we introduce two control methods, i.e., region and shape control, to meet customized requirements.
% To achieve desired surface details and geometric structures, we introduce two control methods: region and shape control. Moreover, we propose using weighted blending for prompt tokens to mitigate semantic coupling issues in SDS, achieving high-quality brush generation. 
The generated VDMs are directly compatible with mainstream modeling software, enabling various applications such as mesh stylization and interactive modeling.
% Additionally, VDM brushes often involve sub-object level structure, which can cause semantic coupling issues in SDS. We propose that applying CFG-weighted blending to the tokens in the prompt can effectively mitigate it, resulting in high-quality sub-object structure generation.
% We expect Text2VDM to more effectively translate recent AI achievements in text-to-image generation into the 3D asset creation workflows used by artists.

\noindent\textbf{Limitations and Future Work}. While our framework can generate high-quality VDM brushes, they may encounter multiview inconsistencies, a common issue introduced by SDS. 
View-consistent diffusion models like MV2MV~\cite{MV2MV} may be helpful to further handle this. Additionally, errors of the T2I models may guide wrong 3D generation, such as the wrong number of fingers in the generated skeleton hand. 
VDMs demonstrate that complex 3D models can be efficiently created using diverse reusable sculpting brushes. Future exploration of 3D generation through the assembly of modular components with similar design principles holds promising research value.

% \vspace{-0.1 cm}
\section*{Acknowledgments}
\label{sec:acknowledgement}
This work was partially supported by the Guangdong Provincial Key Lab of Integrated Communication, Sensing and Computation for Ubiquitous Internet of Things \#2023B1212010007.

{
    \small
    \bibliographystyle{ieeenat_fullname}
    \bibliography{references}
}
\appendix
\maketitlesupplementary
%%%%%%%%% BODY TEXT - ENTER YOUR RESPONSE BELOW
In this supplementary material, we provide additional details and results that are not included in the main paper due to the space limit.

\section{Implementation details}
\begin{figure}[tbh]
\centering
\includegraphics[width=0.5\textwidth]{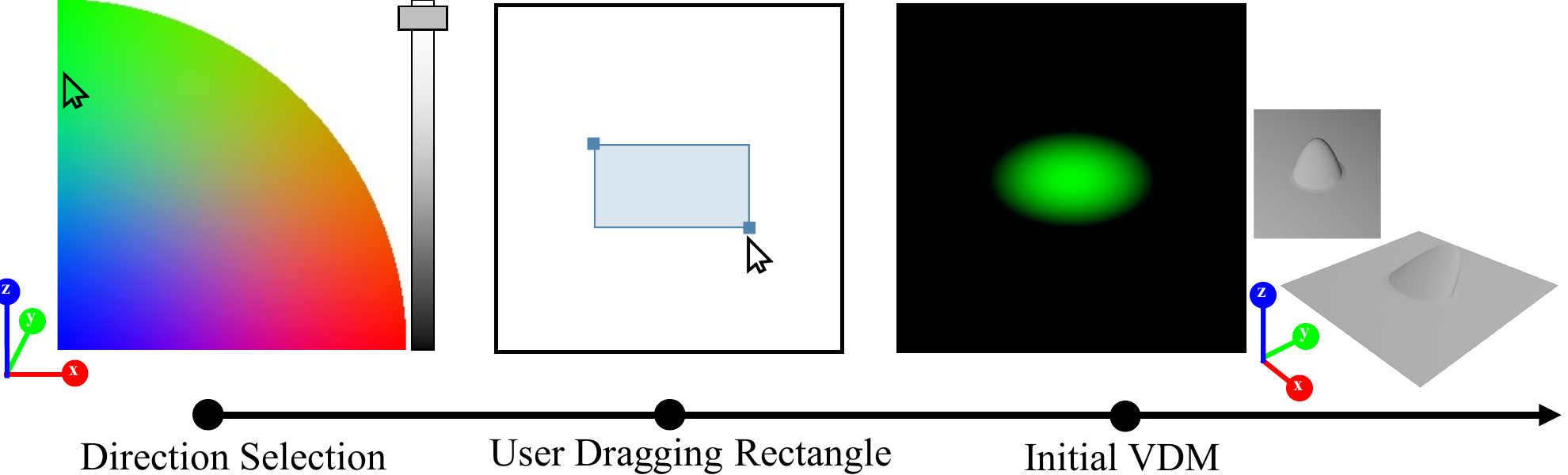}
\caption{
    \textbf{Interface design.} 
}
\label{Fig: Interface Design}
\end{figure}

% 我们的框架提供了一个简单直观的交互界面来辅助用户绘制user-specified VDM来初始化base mesh. 每一轮迭代 we 在我们设置的相机分布中随机render 四张不同视角的 normal images at a resolution of 512×512 pixels.我们的相机部分是。。。。。. We then feed them into the Stable Diffusion 2.1 to caculate SDS loss. 在烘焙过程中，我们把最终mesh的每个顶点的坐标值写入对应的像素点中，得到一张三通道, 分辨率为512×512的exr文件格式的图片，这个exr格式的图片作为最终的VDM输出. The generation process runs on a single NVIDIA RTX 4090 GPU with 10, 000 iterations per brush, which takes around 40 minutes.
\noindent
\textbf{Shape Control.}
Our framework provides a simple and intuitive interactive interface (\Cref{Fig: Interface Design}) to assist users in drawing user-specified VDMs to initialize the base mesh. 

\noindent
\textbf{Brush Optimization.}
In each iteration, We randomly sample camera poses in multi-view and render $N$ view normal images at a resolution of 512×512 pixels. We sample the elevation angle as $\phi_{\text{elev}} \sim \mathcal{U}\left(0, \frac{\pi}{3}\right)$, and the azimuth angle as $\phi_{\text{azim}} \sim \mathcal{U}(0, 2\pi)$. We set $N = 4$ in our experiment. We then feed them into the Stable Diffusion 2.1 to calculate the score distillation sampling (SDS) loss. The generation process runs on a single NVIDIA RTX 4090 GPU with 10, 000 iterations per brush, which takes around 40 minutes.

\noindent
\textbf{VDM Baking.}
For the VDM baking, we write the coordinate values of each vertex of the final mesh into the corresponding pixels, resulting in a three-channel image with a resolution of 512×512 in EXR file format. This EXR image serves as the final VDM output of the generated brush. 
 
\subsection{Interface Design for Shape Control}

\subsection{Details of Weighted Blending}
\begin{figure}[tbh]
\centering
\includegraphics[width=0.3\textwidth]{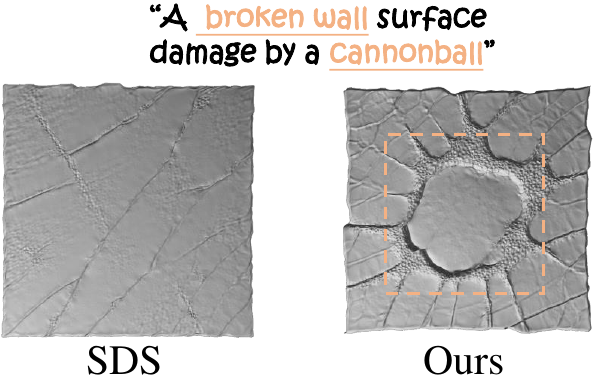}
\caption{
    \textbf{Effect of Semantic Enhancement SDS on surface details generation.} 
}
\label{Fig: surface}
\end{figure}

% 为了获得semantic focus的text embedding来得到更精确的目标分布来缓解原生SDS来的的语义耦和问题，we propose to enhance the semantics of part-related words by applying classifier-free guidance (CFG) weighted blending to the tokens in the prompt. 具体而言, we assign each word in the prompt a CFG weight $s$ and compute the weighted embedding $e_w$ for each word by blending original text embedding $e$ with the empty text embedding $e_{\phi}$ as follows: $e_w = e_{\phi} + s\cdot (e - e_{\phi}) $. By concatenating the weighted embeddings of each word in sequence, we obtain the final semantically focused text embedding。for instance，我们输入prompt：“A horn++ of deer”, 其中++代表对“horn”这个单词进行加权，权重值为1.1^2.正文中所展示的prompt中的黄色下划线单词代表我们在计算text embedding时对该单词进行了权重值为1.21的加权。需要注意的是，这里的CFG权重是计算text embedding时使用的CFG权重，与SDS loss计算过程中使用的CFG guidance scale是分开的。 由于我们使用normal map这种与自然图片差距较大的图片来进行SDS loss计算，我们设置CFG guidance scale 为100。
To obtain semantically focused text embeddings for a more precise target distribution to mitigate the semantic coupling issue in SDS, we propose enhancing the semantics of part-related words by applying weighted blending to the tokens in the prompt. Specifically, we assign each word in the prompt a weight $s$ and compute the weighted embedding $e_w$ for each word by blending the original text embedding $e$ with the empty text embedding $e_{\phi}$ as follows: $e_w = e_{\phi} + s \cdot (e - e_{\phi})$. By concatenating the weighted embeddings of each word in sequence, we obtain the final semantically focused text embedding. For instance, consider the prompt: \textit{``A horn++ of a deer''}, where \texttt{++} indicates that the word \textit{``horn''} is weighted with a weight of $1.1^2$. In the example prompt shown in the main paper, words with yellow underlines represent those weighted with a weight value of $1.21$ during the computation of text embeddings. Notably, the weights used for text embedding computation are separate from the CFG guidance scale applied during the computation of the SDS loss. Our experiment uses a CFG guidance scale of 100. Additionally, our method not only effectively alleviates semantic coupling during geometric structures generation but also enhances specific words in the prompt to produce surface details that better align with the text. For example, \Cref{Fig: surface} demonstrates how enhancing the word "cannonball" results in a better results.

\subsection{Region Control}
\begin{figure}[tbh]
\centering
\includegraphics[width=0.5\textwidth]{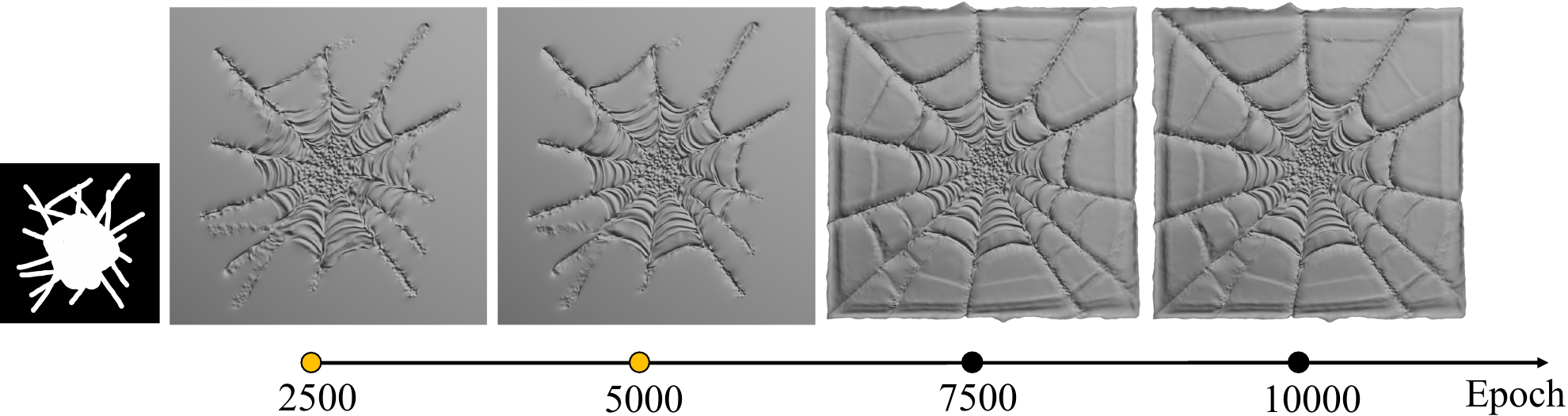}
\caption{
    \textbf{Region control.} 
}
\label{Fig: Region Control}
\end{figure}

% 在我们的框架中，we provide a region mask to restrict mesh deformation to the user-defined region during optimization. By adjusting the activation ratio of the region mask, the final brush effect can effectively match the user’s guidance. 在生成surface details笔刷时，我们activated the region mask for the first half of total iterations as a warm-up stage, 具体控制效果如如中所示。对于geometric structures生成，我们activated the region mask for 整个优化过程。
In our framework, we provide a region mask to restrict mesh deformation to the user-defined region during optimization. By adjusting the activation ratio of the region mask, the final brush effect can effectively match the user’s guidance. When generating surface detail brushes, we activated the region mask during the first half of the total iterations as a warm-up stage, with the specific effects shown in \Cref{Fig: Region Control}. For geometric structures generation, we activated the region mask throughout the entire optimization process to maintain zero values in unused VDM areas.

\section{Details of Experiment Setting}

\subsection{Details of Re-framed Paint-it}
% Paint-it originally uses SDS to optimize a UNet for generating PBR textures. We reframed it to suit our VDM brush generation task. 具体而言，Paint-it原有的架构中使用Unet从一个固定的512×512的高斯噪音图中预测一个九通道输出，每三个通道分别对应diffuse，specular，normal. 我们修改Unet只输出三个通道，对应一个VDM，我们把VDM每个像素的三个值直接应用到对应的mesh顶点进行三轴displacement来得到变形后的mesh，之后我们同样渲染出normal map来计算SDS loss，为了保证mesh的质量，我们添加了促进网格平滑的regulization，如laplacian，edge，normal consistency。
Paint-it originally uses SDS to optimize a UNet for generating PBR textures. We reframed it to suit our VDM brush generation task. Specifically, in the original Paint-it architecture, a UNet is used to predict a nine-channel output from a fixed 512×512 Gaussian noise image, where every three channels correspond to diffuse, specular, and normal maps, respectively. We modified the UNet to output only three channels, representing a VDM. Each pixel's three values in the VDM are directly applied to the corresponding mesh vertices for three-axis displacement, resulting in the deformed mesh. We then render the normal map from the deformed mesh to compute the SDS loss. To ensure the quality of the mesh, we introduced regularization terms to facilitate mesh smoothness, such as Laplacian, edge, and normal consistency regularization.

\subsection{Base Mesh Initialization}
\begin{figure}[tbh]
\centering
\includegraphics[width=0.4\textwidth]{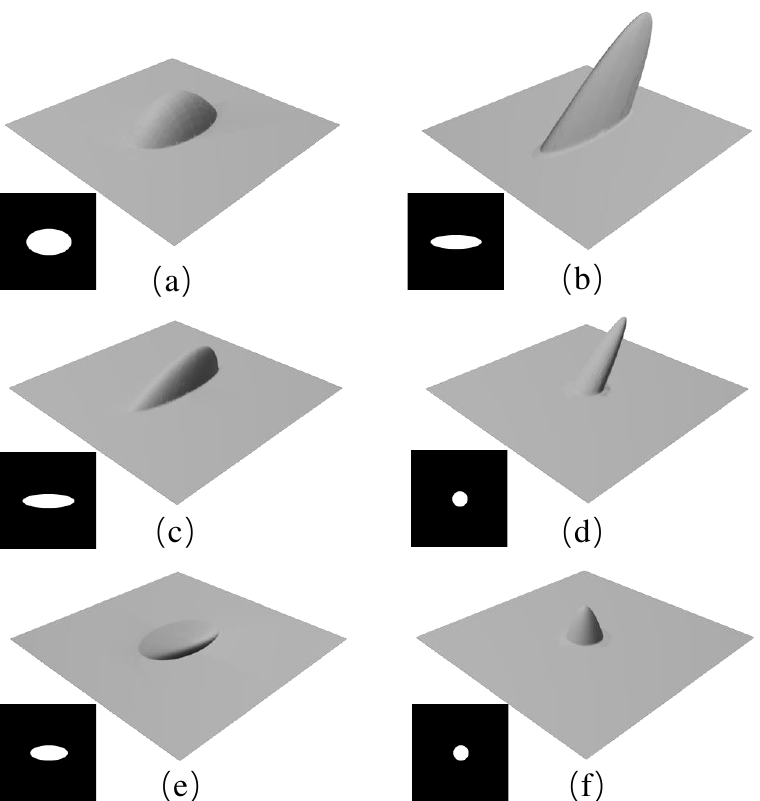}
\caption{
    \textbf{Base mesh initialization.} 
}
\label{Fig: Base Mesh Initialization}
\end{figure}

%在我们生成geometric structures的实验中，为了保证与Text2Mesh,TextDeformer工作的公平比较，我们使用同样的初始base mesh as shown in 图片，具体的prompt与base mesh的对应我们在table1中进行了标注。因为Pint-it是直接生成VDM的方法，我们只使用planar base mesh。
In our experiments for generating geometric structures, to ensure a fair comparison with Text2Mesh and TextDeformer, we used the same initial base mesh and masks as shown in \Cref{Fig: Base Mesh Initialization}. The specific correspondence between prompts and base meshes is detailed in \Cref{tab:table1}. Since Paint-it directly generates VDMs, we only used a planar base mesh and do not use masks.

\subsection{Text Prompt of Quantitative Comparison}
As described in Section 4.2 of the main paper, we used 40 text prompts for VDM generation and quantitative evaluation, with 20 focusing on surface details and the remaining 20 on geometric structures. The specific text prompts are detailed in \Cref{tab:table1}.

% A Table
\begin{table}[thb]\centering
    \caption{Text Prompts for VDM Generation.}
    \label{tab:table1}
    \resizebox{0.48\textwidth}{!}{
    \large
    \begin{tabular}{*{10}{c}}
       %\toprule
       VDM Type &  Text Prompt\\
        \midrule
        Surface Details 
         &  A blanket surface with many flowers \\
         &  A brick wall surface with neat brick \\
         &  A broken brick wall surface with many cracks \\
         &  A broken glass surface like spider web \\
         &  A broken wall surface damaged by a cannonball \\
         &  A broken wall surface with many cracks \\
         &  A cloth surface with a rose pattern \\
         &  A cloth surface with a sunflower pattern \\
         &  A torn cloth surface \\
         &  An arid land surface \\
         &  A stone surface with many cracks \\
         &  A broken stone surface with many cracks \\
         &  A dragon skin surface with many scales \\
         &  A lion fur surface \\
         &  A new wooden floor \\
         &  An aged wooden surface \\
         &  A rusty metal surface with many cracks \\
         &  A surface of rusty metal \\
         &  A skin surface with a scar mended by needle and thread \\
         &   A skin surface with terrible wounds \\
        \midrule
         Geometric Structures  
         &  A lip of human (a)\\
         &  A human spine (a)\\
         &  A mouth of a monster (a)\\
         &  A skeleton hand (a)\\
         &  A tortoiseshell (a)\\
         &  A round snail shell (a)\\
         &  An eye of a monster (a)\\
         &  A dragon wing (b)\\
         &  An angel wing (b) \\
         &  An ear of the devil (b)\\
         &  A dorsal fin of a fish (c)\\
         &  A horn of a dragon (d)\\
         &  A goat horn (d)\\
         &  A horn of the devil (d)\\
         &  A horn of a deer (d)\\
         &  An octopus tentacle (d)\\
         &  An ox horn (d)\\
         &  A royal pauldron (e)\\
         &  A beard of a man (e)\\
         &  A human ear (f)\\
        %\bottomrule
    \end{tabular}
    }
\end{table}

\begin{figure*}[h!tb]
\centering
\includegraphics[width=1\textwidth]{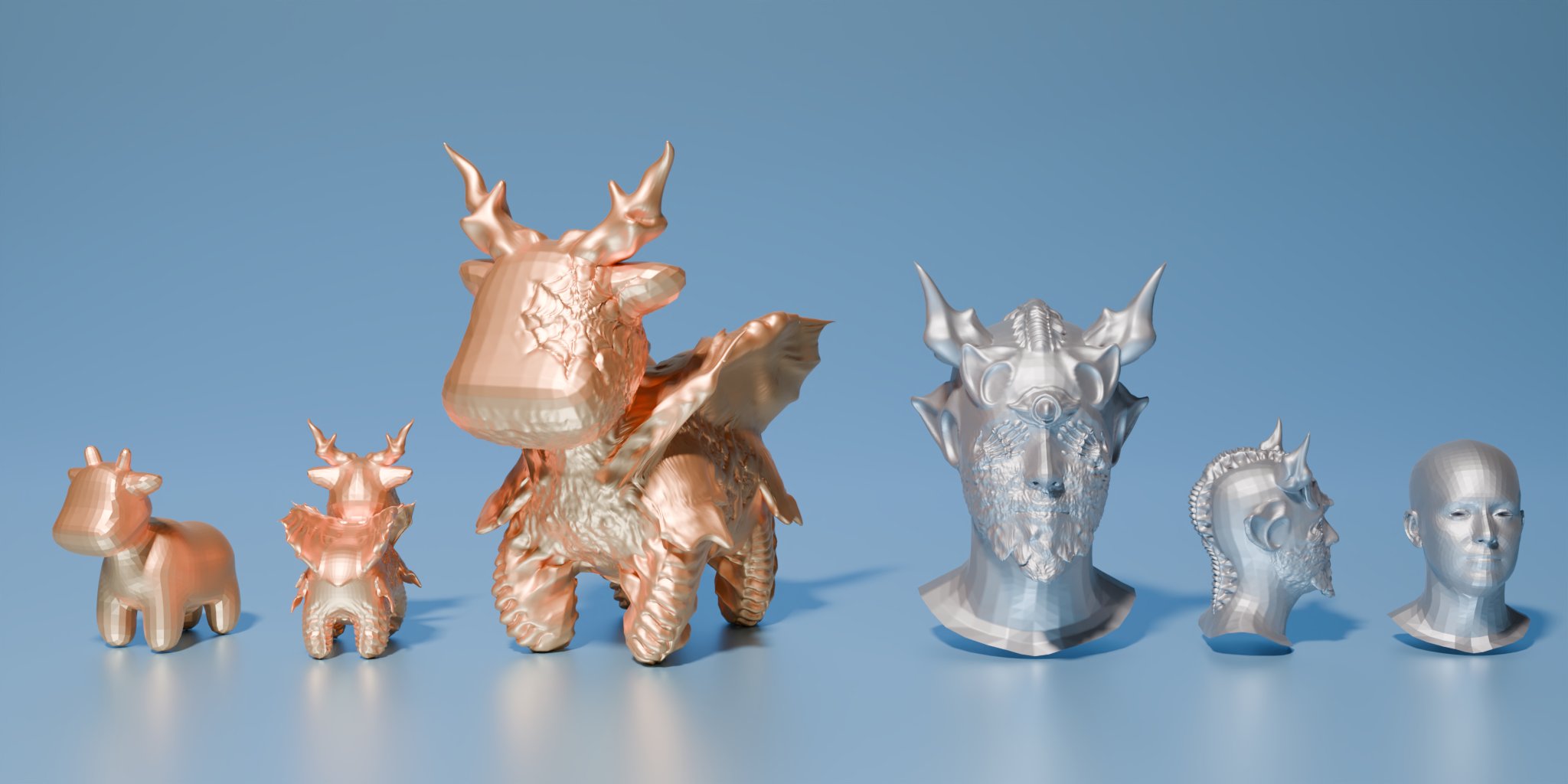}
\caption{
    \textbf{More application examples of interactive modeling.} 
}
\label{Fig: Application}
\end{figure*}

\section{Additional Results}

\subsection{More Examples of Semantic Coupling}
\begin{figure}[tbh]
\centering
\includegraphics[width=0.5\textwidth]{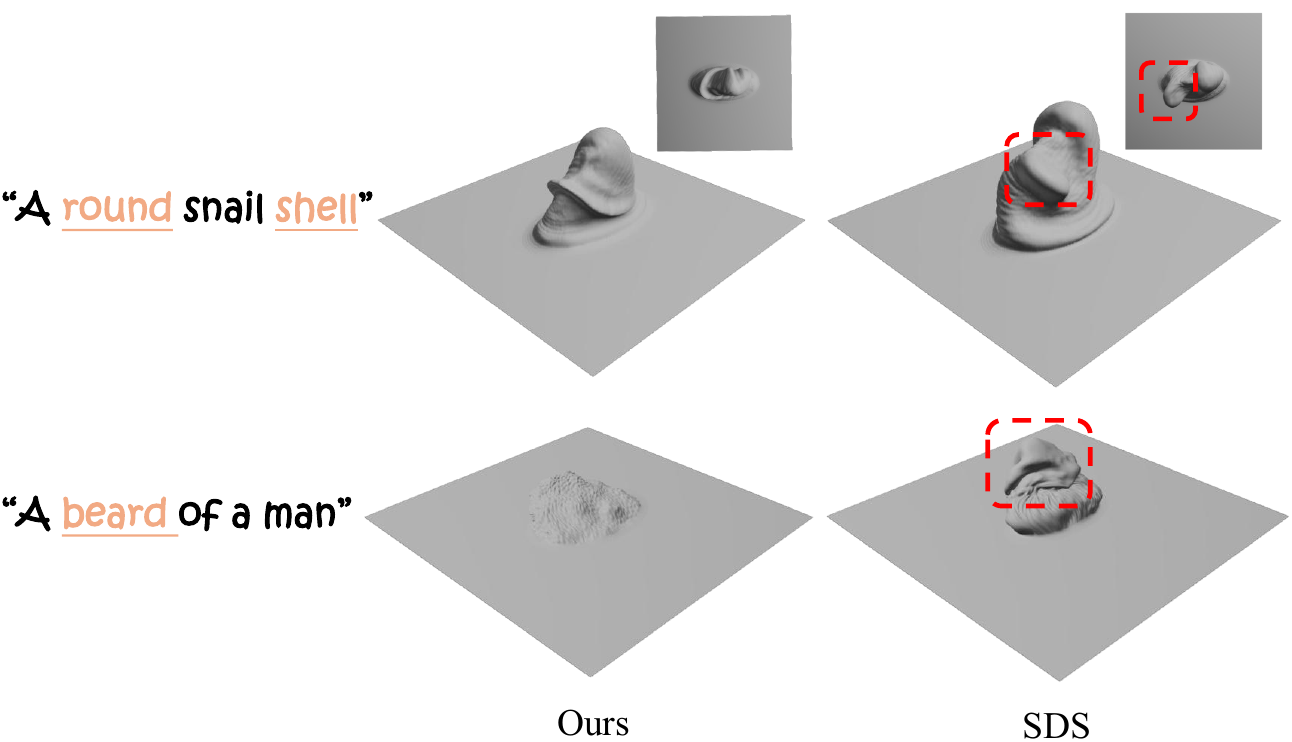}
\caption{
    \textbf{More examples of semantic coupling.} 
}
\label{Fig: More Examples of Semantic Coupling}
\end{figure}

%我们在这小节中展示了更多语义耦合的例子. 如图所示，直接使用SDS进行鹿角的生成会导致生成整个的鹿头，或者在生成胡子时会连带生成鼻子。我们的方法能够显著缓解语义耦生成高质量的sub-object structure.
In this subsection, we present more examples of semantic coupling. As shown in \Cref{Fig: More Examples of Semantic Coupling}, directly using SDS for generating a deer horn leads to the generation of the entire deer head, or generating a beard also results in the generation of the nose. Our method significantly alleviates this semantic coupling issue, enabling the generation of high-quality sub-object structures.

\subsection{More Comparison Results}
\begin{figure*}
\centering
\includegraphics[width=0.65\textwidth]{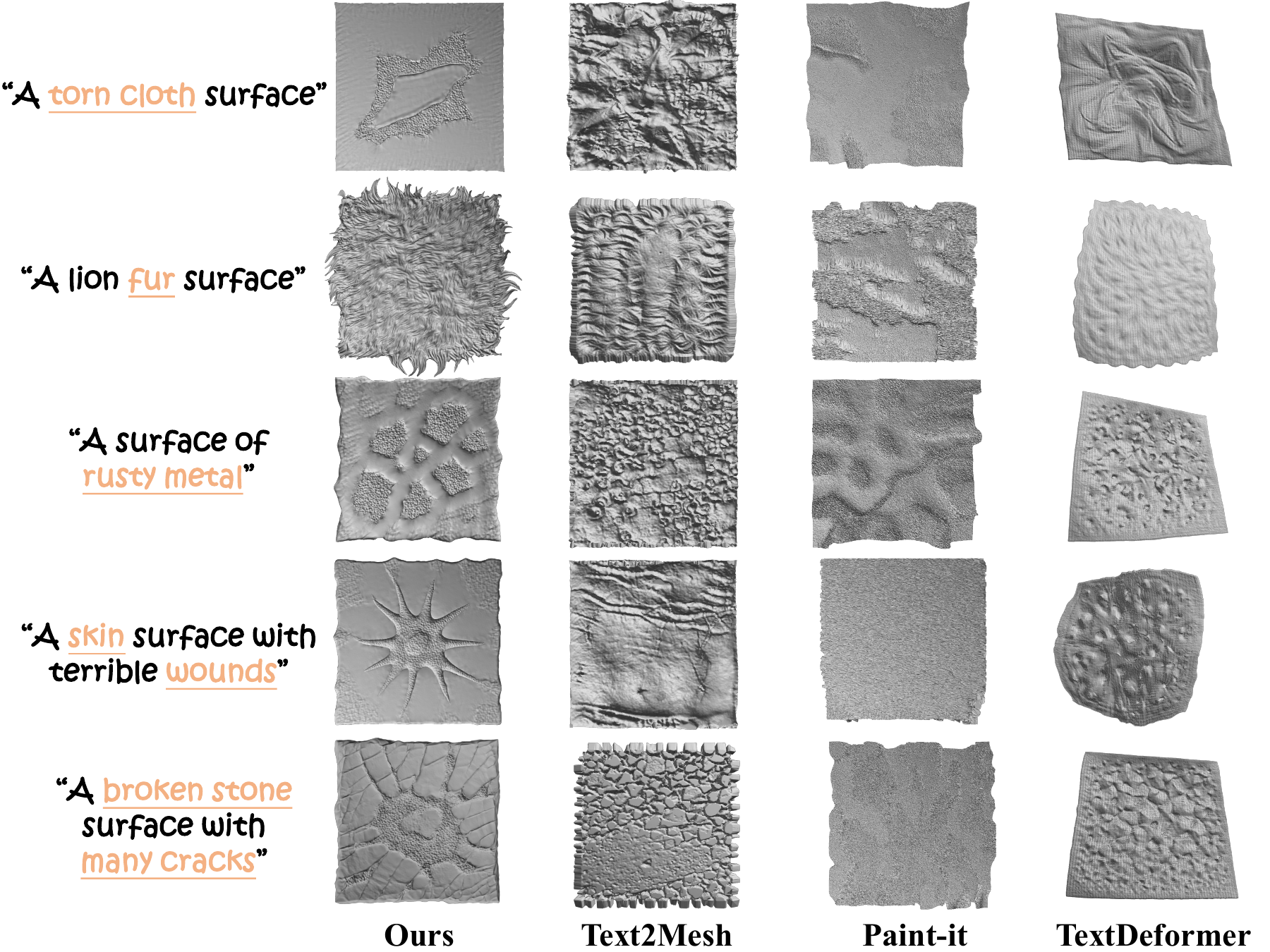}
\caption{
    \textbf{More comparison results of brushes for surface details.} 
}
\label{Fig: surface details}
\end{figure*}

\begin{figure*}
\centering
\includegraphics[width=1\textwidth]{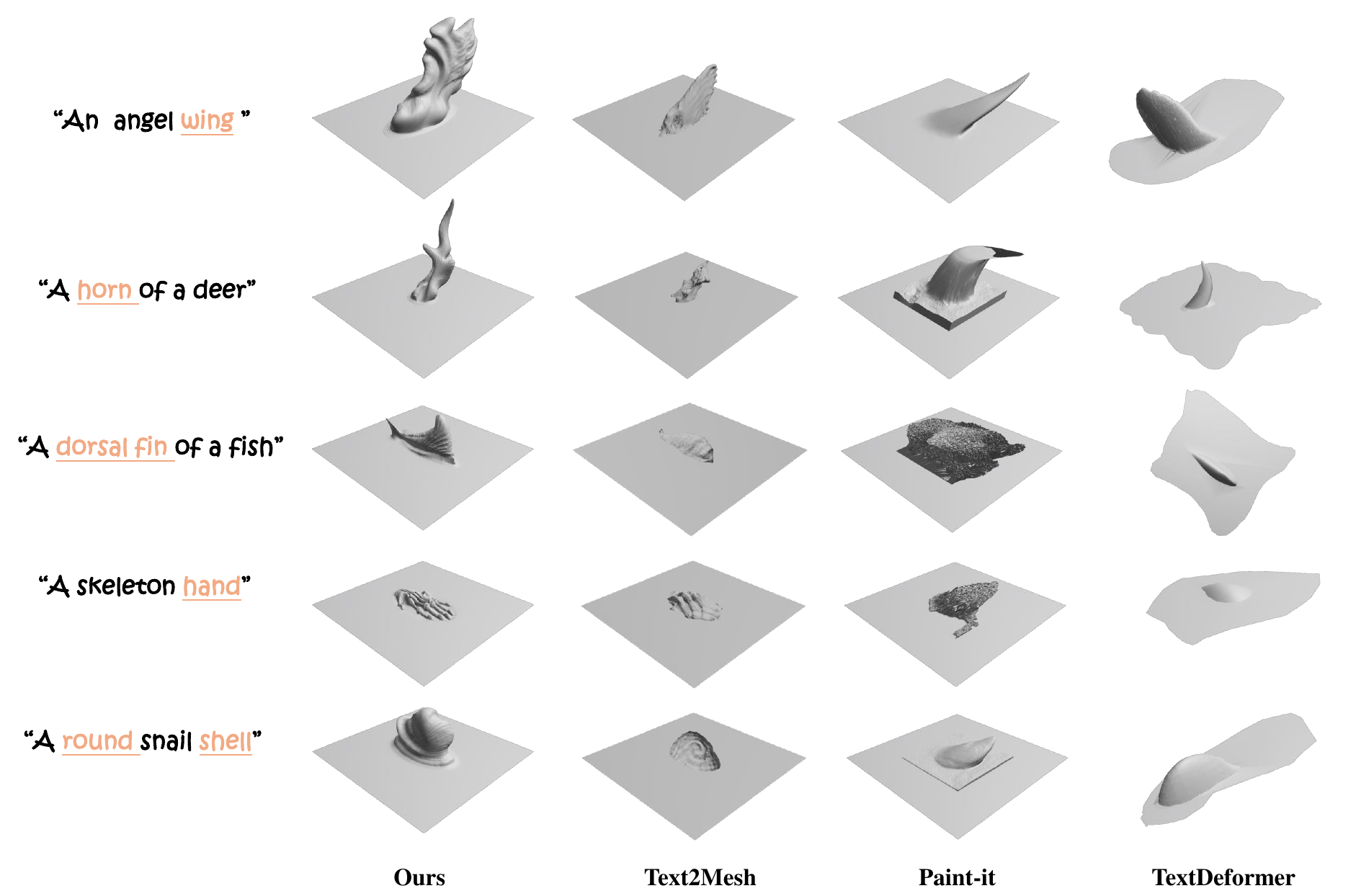}
\caption{
    \textbf{More comparison results of brushes for geometric structures.} 
}
\label{Fig: geometric structures}
\end{figure*}

%我们展示了更多的Qualitative Results for surface details and geometric structures. 我们的方法能够生成更高质量更vivid的结果，as shown in 图片. 
We provide more qualitative results for surface details and geometric structures. Our method can generate higher-quality and more vivid results, as shown in \Cref{Fig: surface details} and \Cref{Fig: geometric structures}.

\begin{figure*}[tbh]
\centering
\includegraphics[width=1\textwidth]{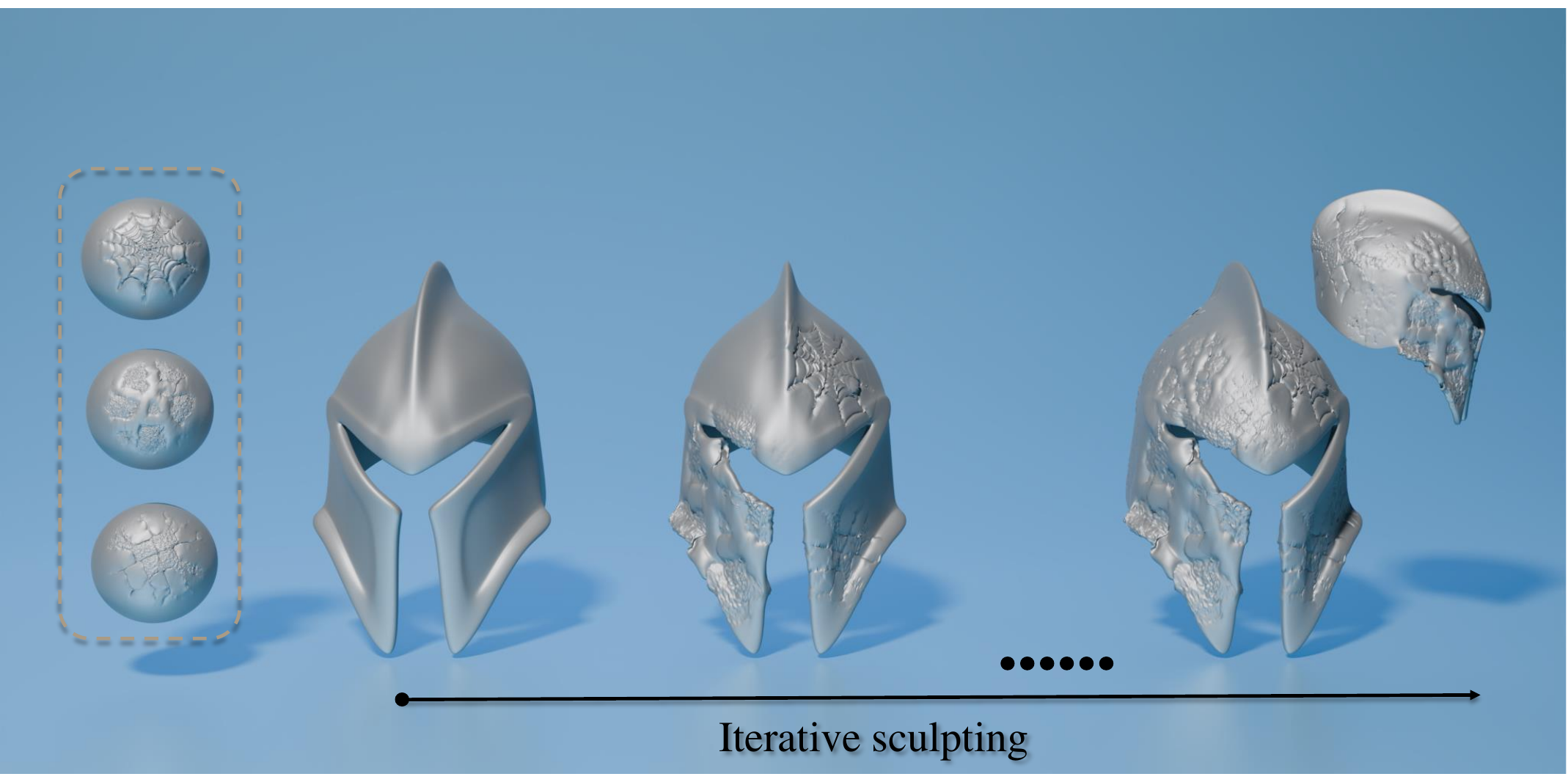}
\caption{
    \textbf{More application examples of mesh stylization on a helmet.} 
}
\label{Fig: Application}
\end{figure*}

\begin{figure*}[tbh]
\centering
\includegraphics[width=1\textwidth]{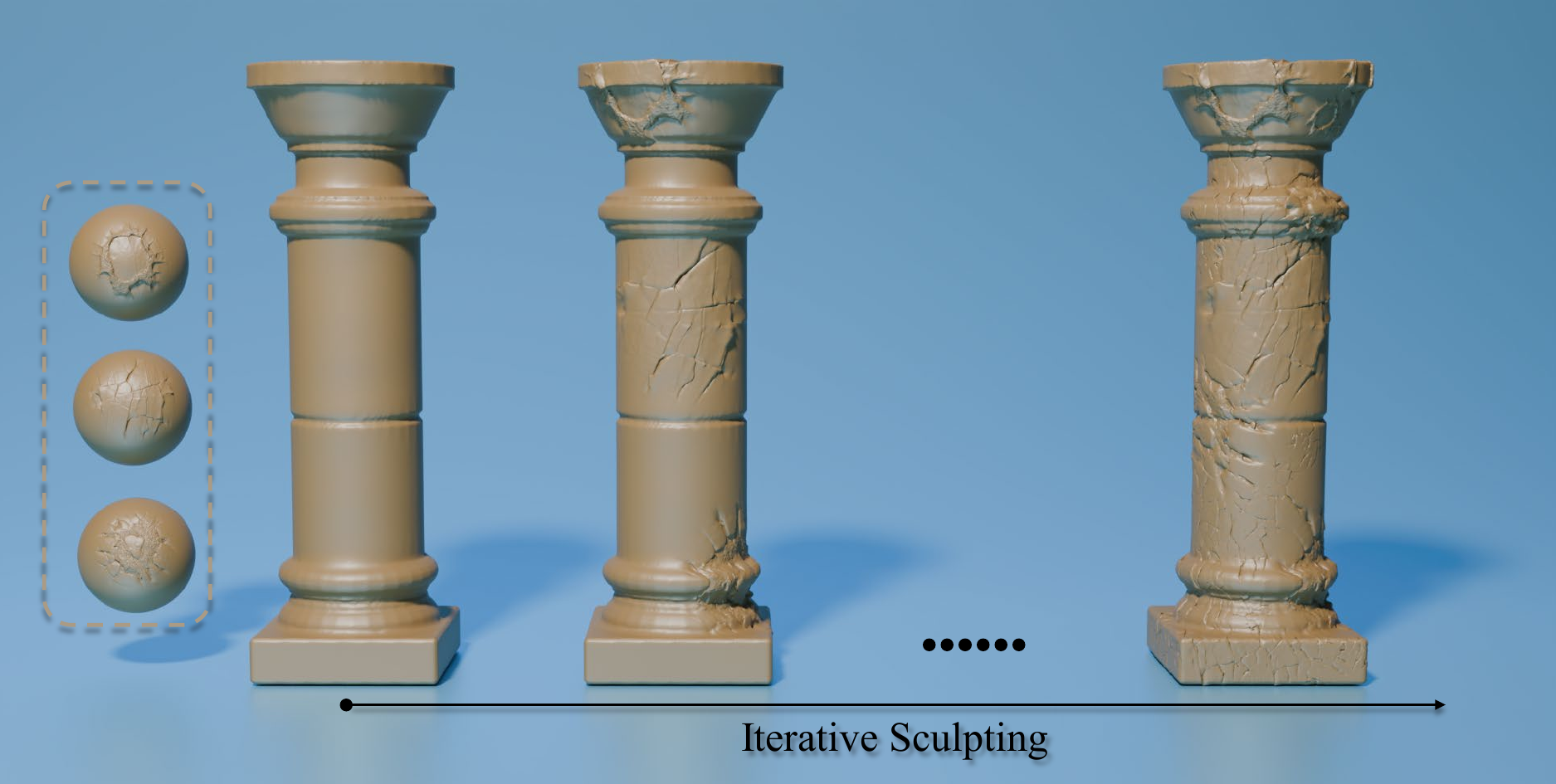}
\caption{
    \textbf{More application examples of mesh stylization on a pillar.} 
}
\label{Fig: Application}
\end{figure*}

\subsection{More Application Examples}
%我们展示了更多使用我们生成的笔刷进行雕刻的例子，We enable users to interactively use a variety of generated brushes to sculpt diverse and expressive models from a plain shape.
We also provide more examples of sculpted models using the generated VDM brushes. We enable users to interactively use a variety of generated brushes to sculpt diverse and expressive models from a plain shape.

\end{document}